\documentclass[11pt,onecolumn,draftcls,peerreview]{IEEEtran}
\newcommand{\numcol}{2}

\usepackage{setspace}
\usepackage{verbatim}

\usepackage{amsmath}
\usepackage{amssymb}
\usepackage{graphicx}
\graphicspath{{eps/}}
\usepackage[noadjust]{cite}
\usepackage{subfigure}
\usepackage{multirow}

\usepackage{enumerate}
\usepackage{ifthen}
\newcommand{\checkOneCol}{\ifthenelse{\numcol = 1}}

\begin{document}

\newcommand{\papertitle}{A Self-Organization Framework for Wireless Ad Hoc Networks as Small Worlds}

%%%%%%%%%%%%%%%%%%%%%%%%%%%%%%%%%%%%%%%%%%%%%%%%%%%%%%%%%%%%%%%%%%%%%%%%%%%%%%%%

\title{\papertitle}

\author{\authorblockN{Abhik Banerjee\authorrefmark{1}, Rachit Agarwal\authorrefmark{2}, Vincent Gauthier\authorrefmark{2}, Chai Kiat Yeo\authorrefmark{1}, Hossam Afifi\authorrefmark{2} and Bu Sung Lee\authorrefmark{1} \\}
\authorblockA{\authorrefmark{1} CeMNet, School of Computer Engineering, Nanyang Technological University, Singapore\\}
\authorblockA{Email: \{abhi0018, asckyeo, ebslee\}@ntu.edu.sg\\}
\authorblockA{\authorrefmark{2}Lab. CNRS SAMOVAR UMR 5157, Telecom Sud Paris, Evry, France\\}
\authorblockA{Email: {\{rachit.agarwal, vincent.gauthier, hossam.afifi\}}@telecom-sudparis.eu}
}

\maketitle

%%%%%%%%%%%%%%%%%%%%%%%%%%%%%%%%%%%%%%%%%%%%%%%%%%%%%%%%%%%%%%%%%%%%%%%%%%%%%%%%
\begin{abstract}
Motivated by the benefits of small world networks, we propose a self-organization framework for wireless ad hoc networks. We investigate the use of directional beamforming for creating long-range short cuts between nodes. Using simulation results for randomized beamforming as a guideline, we identify crucial design issues for algorithm design. Our results show that, while significant path length reduction is achievable, this is accompanied by the problem of asymmetric paths between nodes. Subsequently, we propose a distributed algorithm for small world creation that achieves path length reduction while maintaining connectivity. We define a new centrality measure that estimates the structural importance of nodes based on traffic flow in the network, which is used to identify the optimum nodes for beamforming. We show, using simulations, that this leads to significant reduction in path length while maintaining connectivity.
\end{abstract}

\section{Introduction}\label{sec:Intro}
Given the diversity and scale of deployment of future wireless networks, self-organization is important for ensuring scalability and reliability. As wireless ad hoc networks typically suffer from issues of reliability and scalability, it is necessary to have algorithm designs that are not just distributed but also operate with purely local knowledge \cite{DresslerSONAdHoc}. Prehofer and Bettstetter outlined a set of design principles for self-organization in wireless networks in \cite{PrehoferSlfOrg}. 

\begin{figure}[tb]
    \centering
    \mbox{\subfigure[Rewiring with increasing value of $p$]{
    \includegraphics[width=3in]{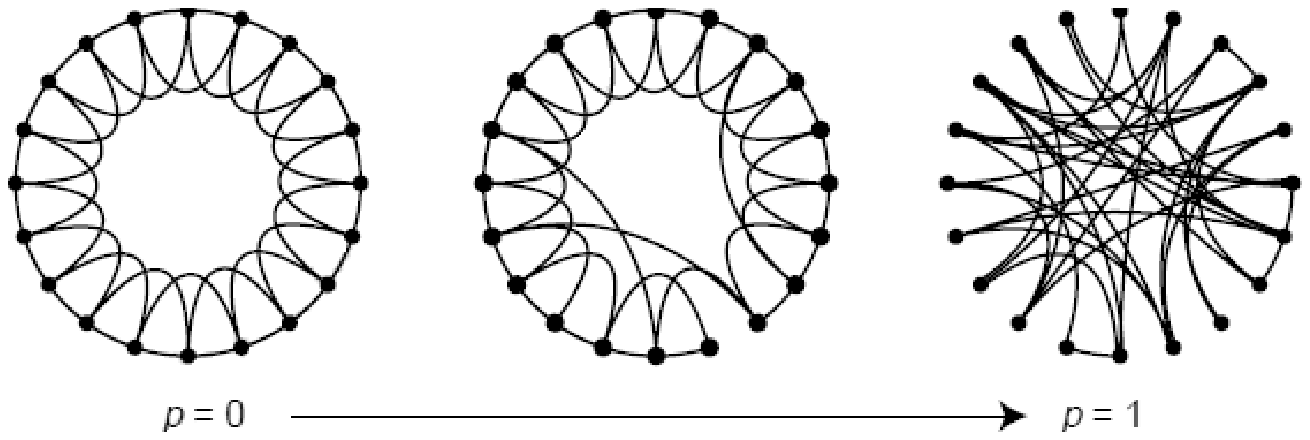}
    \label{subfig:watts_swrew}
    }\quad
    \subfigure[Reduction in APL and CC with increase in $p$.]{
    \includegraphics[width=3in]{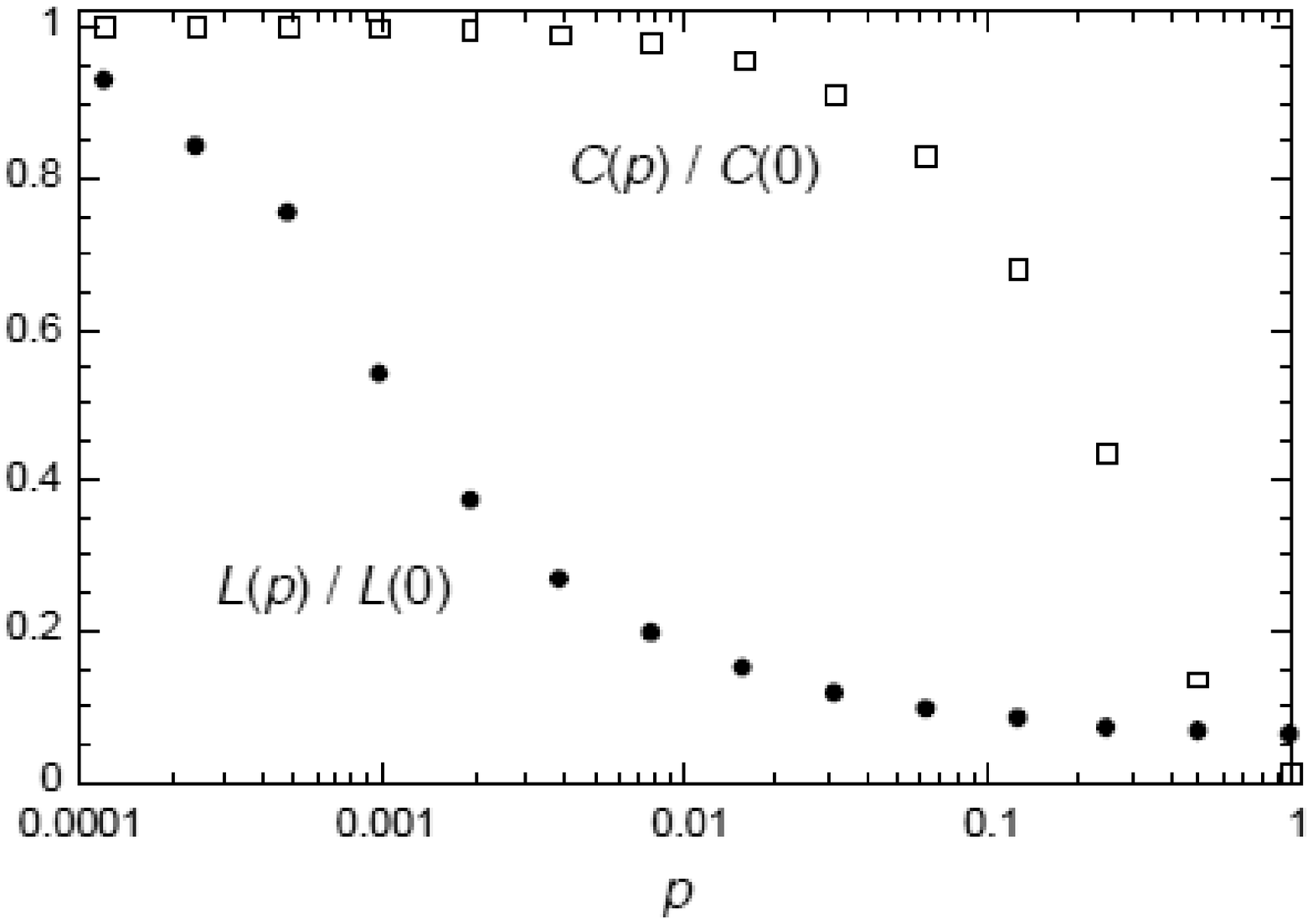}
    \label{subfig:watts_swgraph}
    }}
    \caption{Small World properties achieved by random rewiring in a regular network \cite{WattsStrogSWN}.}
    \label{fig:watts_swgraph}
\end{figure}

An attractive model of reorganization to ensure performance guarantees in large scale networks is the small world network. The small world phenomenon was initially studied in the context of social networks by Stanley Milgram \cite{MilgramSW} who observed that the average separation between a source and target typically lies between five and six, subsequently termed and popularised as "six degrees of separation". It was shown by Watts \& Strogatz \cite{WattsStrogSWN} that by rewiring a small set of links in a regular network, small world behaviour could be realized, characterized by short average path length and high clustering coefficient.  Here, the average path length (APL) refers to the mean separation over all possible node pairs in the network. The clustering coefficient (CC) measures the fraction of a node's neighbourhood that are also each other's neighbours \cite{HelmySWWi}. The model proposed in \cite{WattsStrogSWN} is shown in Fig. \ref{fig:watts_swgraph} where all edges in a regular network are rewired with a probability $p$. As seen in Fig. \ref{subfig:watts_swgraph}, even rewiring only $1 \%$ of edges results in a reduction in APL by almost $80 \%$ while the CC hardly changes. From the perspective of network design, smaller values of APL imply performance guarantees for data delivery since paths between all node pairs are bounded. This aspect of small world networks was discussed in \cite{MartelAnalSWModel}, where it was showed that the diameter of such a network grows with the logarithm of the network size. Kleinberg \cite{KleinbergSWN} showed that a decentralized routing algorithm can result in short paths for a grid based model. Further, design of a decentralized routing protocol for a small world network can be designed with low state information at individual nodes \cite{FraigniaudEclect}. Coupling of short APL with high values of CC implies high connectivity in the network, thereby making it robust to changes. All these characteristics underscore the attractiveness of small worlds as a self-organization design goal.

The salient features of small world networks are highly desirable from the point of view of the diversity of deployments of future wireless networks. Networks such as pocket switched networks (PSNs) and vehicular ad hoc networks (VANETs) are likely to be characterized by an increase in dynamicity not just because of the scale and diversity of the wireless devices in operation but also due to the range of applications they are likely to cater to \cite{HuiPSNSummary,KrishnanVANETApps,BaiCharAutApps}. It is therefore necessary for the design of network architecture that can scale to a variety of traffic requirements that range from ensuring service guarantees for multimedia traffic such as voice and video \cite{AsefiVidVANET,SardariRtless,LinSocContDiff} to critical applications such as safety messages in VANETs which require high reliability \cite{ChiasseriniSafetyMsg,KhabazianPerfMsg,BussonAnalIVC}. Moreover, different types of networks such as PSNs and VANETs are likely to coexist rather than being exclusive, thereby necessitating the need for a generic design that ensures performance guarantees irrespective of specific application requirements. It was observed by Gupta and Kumar \cite{GuptaKumar} that the upper bound on the network throughput scales inversely with the APL. Additionally, due to the inherent ad-hoc nature of such networks, it is necessary to look at conditions that allow for network protocol design that not only incurs low overhead but can also be implemented with minimal state information. Such constraints make an ideal case for small world networks as they are characterized by bounded path lengths between all nodes. As a result, routing protocols which involve low state information and incur low overheads \cite{KleinbergSWN,FraigniaudEclect} can be designed, thereby alleviating issues with increasing overhead costs in existing ad hoc routing protocols \cite{ZhouTraffOvhd,SofraLnkLfTm} which in turn lead to greater energy consumption. Further, the high clustering coefficient of small world networks implies high connectivity thereby ensuring reliability.

However, it was noted by Helmy in \cite{HelmySWWi} that owing to the fact that wireless networks are spatial graphs rather than relational, short cut links cannot be completely random as in \cite{WattsStrogSWN}. Rather, the possibility of creating a short cut link between two nodes is determined by the distance between them and the radio transmission range. Subsequent research has looked to identifying different strategies for short cut creation between nodes. Designs proposed in the existing literature \cite{ChitraWired} - \cite{DixitSOCellWiNw}, though, make use of additional infrastructure, such as wires, additional radios and high capacity nodes, for creation of short cuts. Such a design limits them to specific scenarios when the corresponding infrastructure is present. A second limitation of the literature discussed above is that all of them consider at least some knowledge of the global network. Acquiring such information incurs additional overheads, thereby impacting the network performance.

The focus of this paper is to design a framework for wireless networks to self-organize as small worlds using only locally available information. We study how small world behavior can be realized in a wireless network by the use of directional beam forming at the nodes. Our primary motivation for using directional antennas primarily stems from the fact that they can be used to transmit over longer transmission ranges than omnidirectional antennas while using the same transmission power. This implies that short cuts can be created between nodes without the need for additional infrastructure. This distinguishes us from existing literature that focus on addition of new links between nodes. In our case, however, existing omnidirectional links are rewired as long range directional ones. Another unique property of using directional antennas is that the shape of the beam implies that long range links can be created with more than just one node. Thus, the fraction of long range links in the network is actually higher than the number of nodes beamforming. Further, recent advances in directional antenna design has thrown open the possibilities of large scale deployment in wireless networks, including wireless sensor networks \cite{GiorgettiLowCostDA,KakoyiannisDA}. Therefore, a design considering directional antennas can adapt to a wide range of network scenarios.

We are interested in studying the issues surrounding the use of directional beamforming for small world creation in a wireless ad hoc network and propose ways to achieve an optimal design. We consider a connected network for which the primary objective is to maximize reduction in path length without any loss in connectivity. To our knowledge, the only other work where the use of directional antennas was mentioned for the purpose of creating short cuts was in \cite{VermaSWWMN}. However, like other existing papers, the proposed model considers addition of links using multiple radios.

In order to evaluate the benefits of directional beamforming for realizing small world behavior, we first do a simulation based analysis of the potential benefits and challenges \footnote{The results presented in this part of the paper were earlier reported in \cite{BanerjeeWFBCCNet}}. We consider a setup in which a fraction of nodes use long range beams in randomly chosen directions. Our results show that significant reduction can be achieved in the average path length of the network. However, the path length improvements are accompanied by a high fraction of paths being asymmetric, which results from the directional nature of links \cite{LiAsymConn,YuConnCov}. Algorithm design for small world creation using directional beamforming, therefore, needs to strike a balance between path length reduction and loss in connectivity.

Motivated by the benefits achievable using randomized beamforming, we shift our focus to distributed algorithm design for small world creation using directional beamforming. Central to our design is a new measure of centrality defined in this paper that allows distributed estimation of the structural importance of nodes in the network. We define Wireless Flow Betweenness (WFB) which gives an accurate estimate of the Flow Betweenness Centrality (FBC) \cite{FreemanFBC}. The proposed measure enhances the earlier measure defined in \cite{BanerjeeWFBCCNet} by identifying key redundancies. The key aspect of WFB is that it can be computed in a completely distributed manner by exploiting the wireless broadcast advantage (WBA) \cite{WieselthierEgBcast} for information regarding traffic flow, thereby incurring negligible overheads. To our knowledge, the only other measure of centrality designed from the perspective of distributed implementation in wireless ad hoc networks is the \emph{Aggregated Weight N-hop Ranking (AWeNoR)} proposed by the authors in \cite{MaglarasPathBased}. However, though it does away with the need for global network information, it still requires explicit multi-hop information. 

We propose a distributed algorithm that makes use of WFB values computed at individual nodes to decide on their beamforming behavior. Using simulations, we show that this results in greater path length reduction than randomized beamforming with negligible effect on connectivity. Compared to randomized beamforming, the beamforming decision adapts to the structure of the network, resulting in better performance. We note here that the discussion in this paper does not concern itself with the MAC and routing aspects of directional antennas, which have received substantial research attention in the past. Any of the existing approaches can be used in tandem with our design. More details can be found in \cite{ChoudhuryDARting,ChoudhuryDAMAC,RamanathanDASoln} and the references therein.

In summary, the primary contributions of this paper are:
\begin{enumerate}[a)]
\item A simulation based analysis of using randomized directional beamforming for small world creation in wireless ad hoc networks. We identify the achievable benefits and challenges.
\item A new measure of betweenness centrality, defined specifically in the context of wireless ad hoc networks, that can be computed individually at nodes to estimate their structural importance in the network. The proposed measure is distributed and lightweight in design. 
\item Distributed algorithm design for deterministic creation of directional beams for small world creation. The proposed algorithm makes use of the betweenness centrality defined earlier to enable decision making at nodes.
\end{enumerate}

The rest of the paper is organized as follows. The next section provides the details of a simulation based analysis of randomized beamforming. Section \ref{sec:dist-cent} introduces the Wireless Flow Betweenness (WFB). A centralized scheme for choosing optimal nodes for beamforming is shown to give promising results for small world creation. Subsequently, a distributed algorithm for small world creation using beamforming  is proposed and evaluated using simulations in section \ref{sec:swn-wfb-algo}. In section \ref{sec:rel-work}, a comparative study is done to suitably position our work with respect to existing literature on self-organization and small worlds in wireless networks. Finally, we conclude the paper in section \ref{sec:concl}.
% Add challenges involved, summary of second parts

\section{Small World Wireless Networks using Directional Antennas}\label{sec:swnda-anal}
In this section, we do a simulation based analysis of using directional antennas for small worlds in wireless networks. We use the results to identify crucial design aspects.

\subsection{Network Model}\label{subsec:nw-model} 
We consider a wireless ad hoc network of $N$ nodes all of which consist of a single beamforming antenna. Initially, all nodes transmit using omnidirectional beams with range $r$. Subsequently, a fraction $p$ of the nodes in the network are randomly chosen which use long range directional beams. Usage of directional antenna by a node can be classified into different categories depending on the modes of tranmission and reception \cite{LiAsymConn}. We consider that when a node creates a long range directional beam, it operates in the mode of directional transmission and omnidirectional reception (DTOR). For the purpose of analysis, we use the sector model for directional antennas \cite{YuConnCov}. We compare the simulation results using the sector model to a more realistic uniform linear array (ULA) antenna \cite{BettstetterRndBeamform} model.

\begin{figure}[tb]
\centering
\mbox{\subfigure[Beam length for $\theta = \frac{\pi}{8}$]{\includegraphics[width=3in]{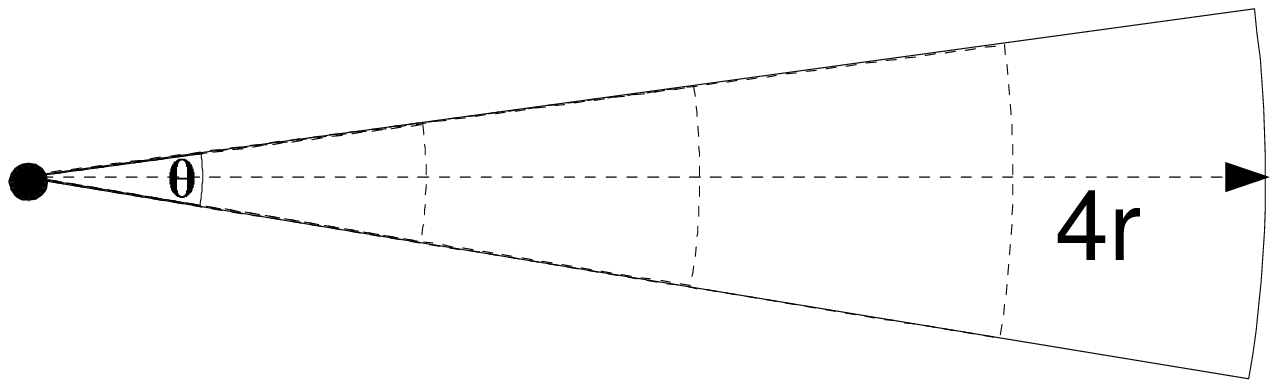}\label{subfig:beamw_pi8}}\quad
\subfigure[Beam length for $\theta = \frac{\pi}{4.5}$]{\includegraphics[width=3in]{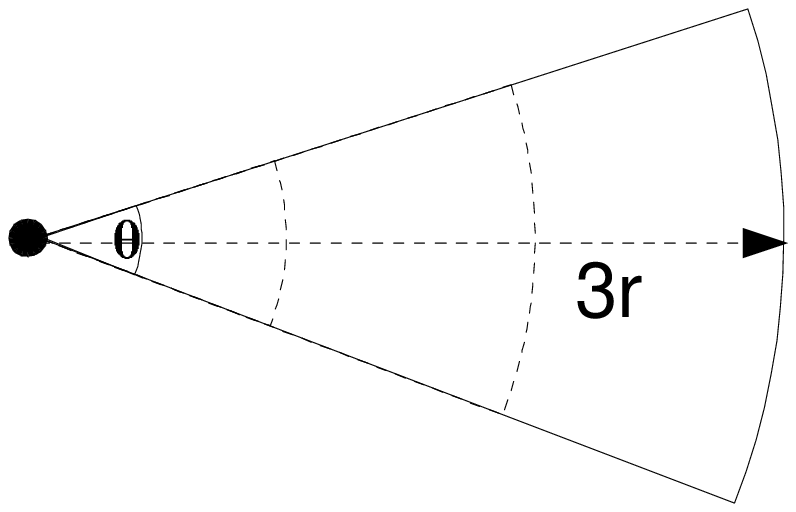}\label{subfig:beamw_pi45} }}
\mbox{\subfigure[Using a narrow beam can result in asymmetric paths]{\includegraphics[width=3in]{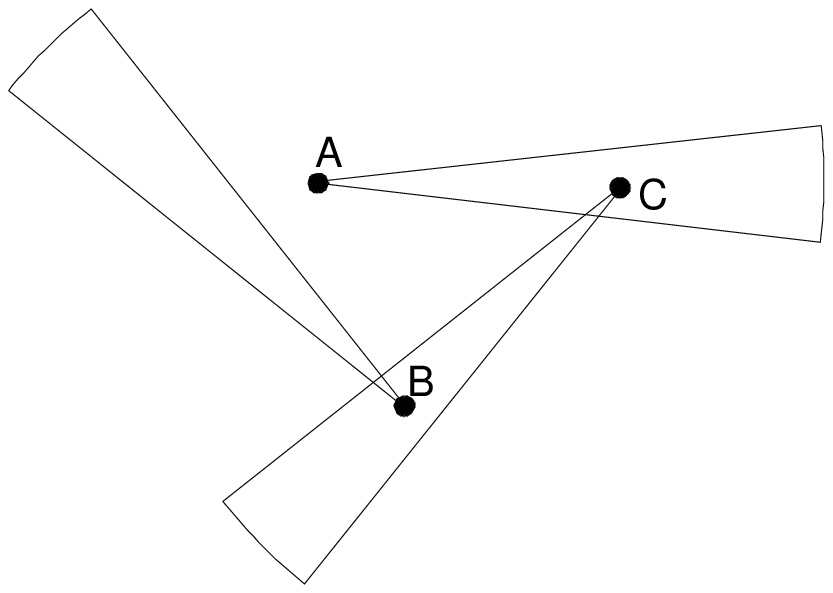}\label{subfig:narrow_beam}}\quad
\subfigure[Using a broader beam can help create a circular path.]{\includegraphics[width=3in]{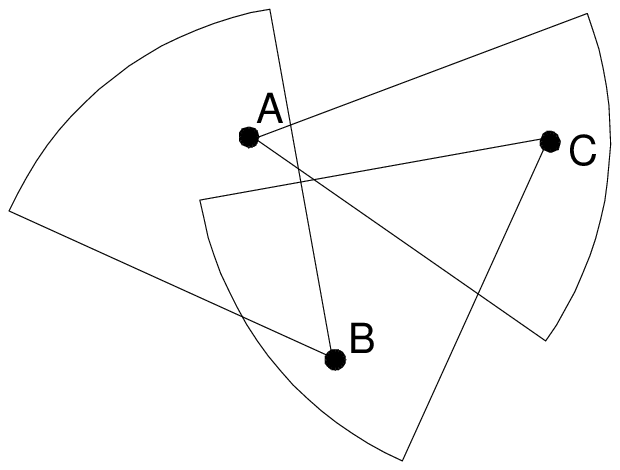}\label{subfig:broad_beam} }}
\caption{Relation between beamwidth and length and the effect on connectivity.}
\label{fig:conn_wid}
\end{figure}

A directional beam is characterized by the beam length, width and the beam direction. For the ULA, a longer beam length can be achieved by increasing the number of antenna elements used \cite{BettstetterRndBeamform}. By keeping the transmission power constant, increasing the number of elements results in a narrower and longer main lobe, while using a single antenna results in an omnidirectional beam. When using the sector model as an abstraction, a constant transmission power implies that the area covered by the beam stays constant. Thus, the beam length depends directly on the width. The resulting beam length $r(\theta)$ for a beam width $\theta$ can be given in terms of the omnidirectional transmission range $r$ as
\begin{equation}\label{eq:lr}
r(\theta) = r \sqrt{\frac{2 \pi}{\theta}}
\end{equation}
The relation between the beamwidth $\theta$ and beamlength is illustrated in Figs. \ref{subfig:beamw_pi8} and \ref{subfig:beamw_pi45}.

%Small world creation may be achieved based on all the above set of parameters characterizing the directional beam as well as the choice of nodes that use long range beams. In this study, we limit the parameters to the beam direction and the choice of nodes. 
The directed nature of the beam leads can lead to the problem of asymmetric paths between nodes, as discussed in \cite{LiAsymConn,YuConnCov}. Ensuring bidirectional traffic flow is supported between a pair of nodes essentially requires the presence of a circular path. Figs. \ref{fig:conn_wid}(c) and \ref{fig:conn_wid}(d) illustrate the effect of beamwidth on the connectivity between two nodes $A$ and $B$. Using a narrow beam initially implies that transmission can only proceed in a single direction from $A$ to $B$. Using a wider beam at $B$ ensures that it can communicate with $A$ as well.
\begin{figure}
	\centering
    \includegraphics[scale=1.0]{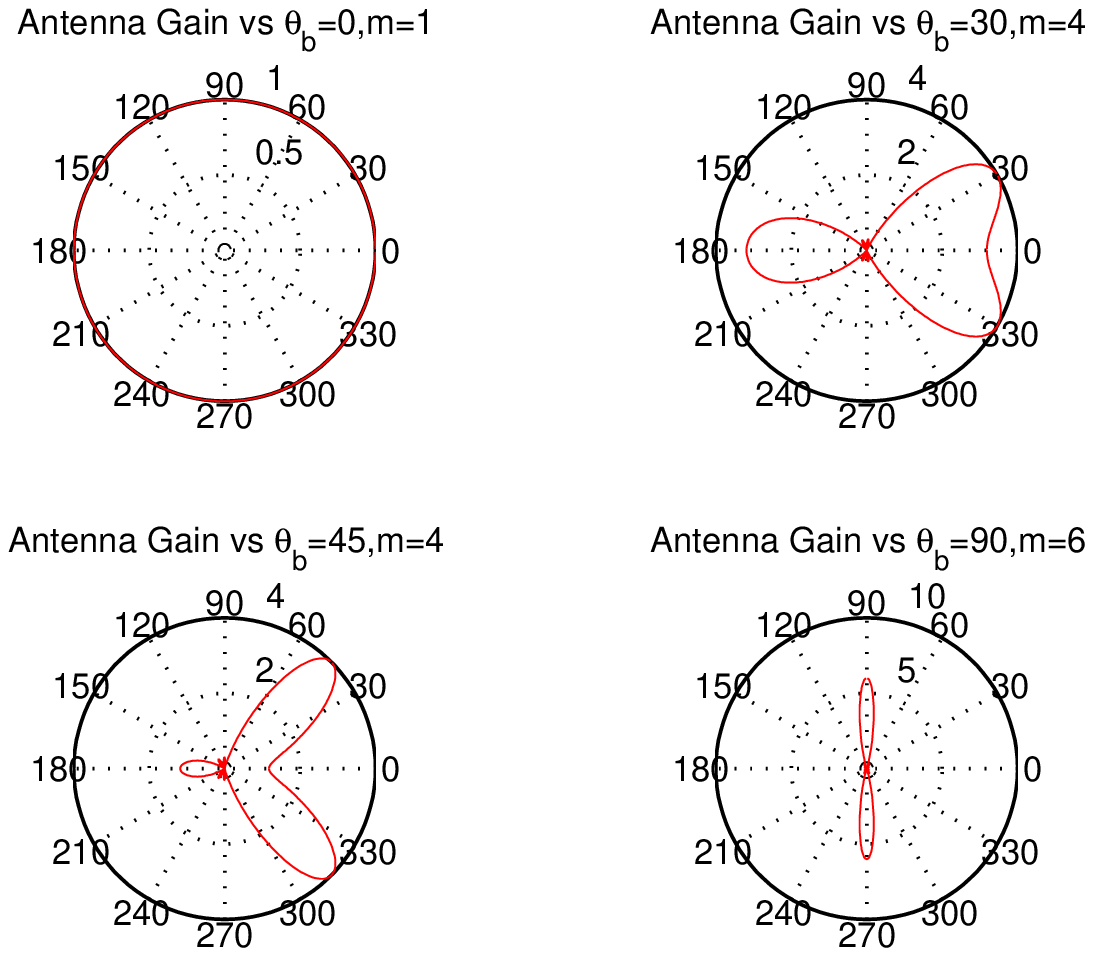}
    \caption{ULA Beam patterns for different values of $\theta_b$ and $m$.}
    \label{fig:beampatt_ula}
\end{figure}

\begin{figure}
\centering
\mbox{\subfigure[Initial Network Setup with Omnidirectional Beams]{\includegraphics[width=3in]{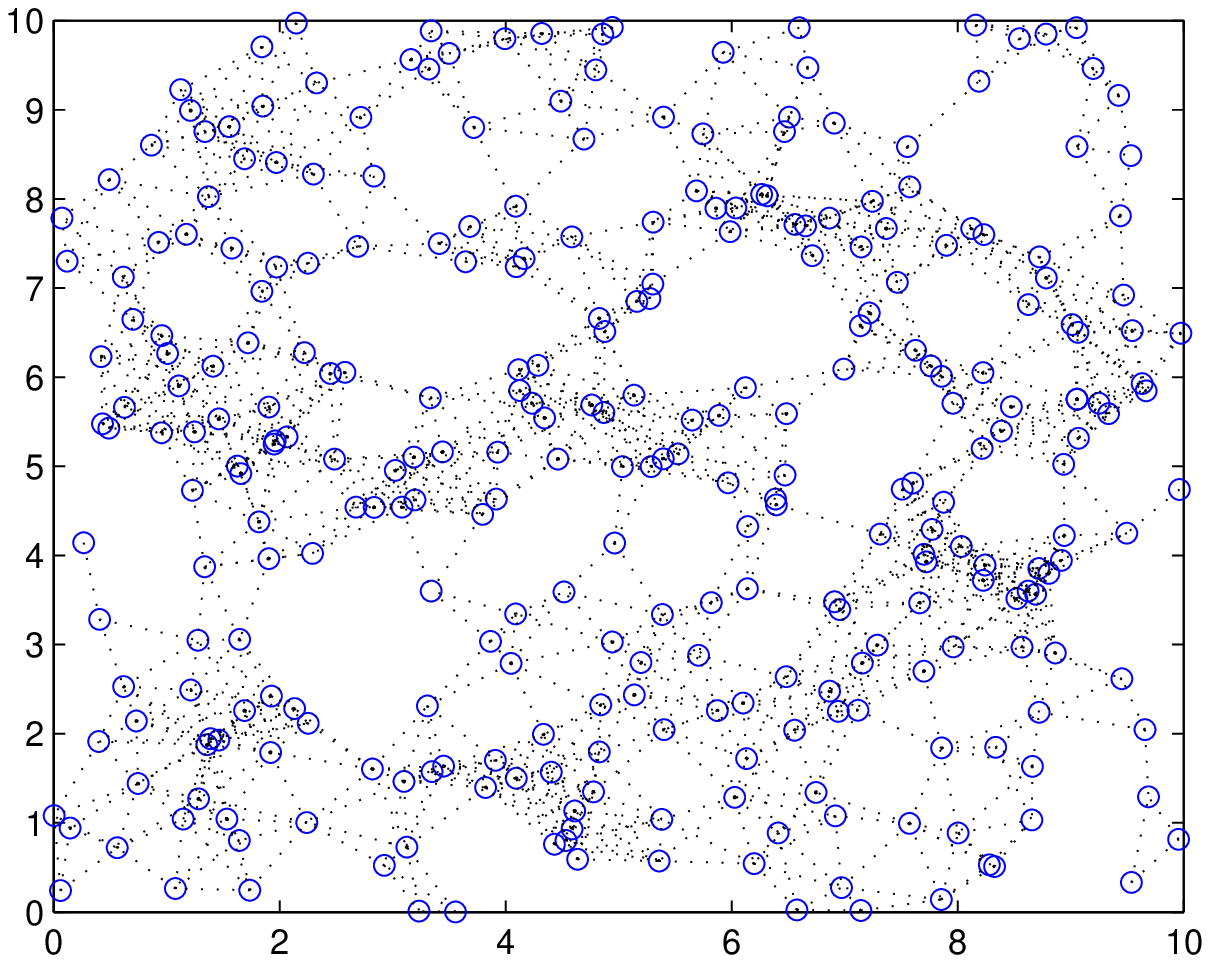}\label{subfig:nwillrnd_initnw}}\quad
\subfigure[Randomized Beams for $p=0.01$]{\includegraphics[width=3in]{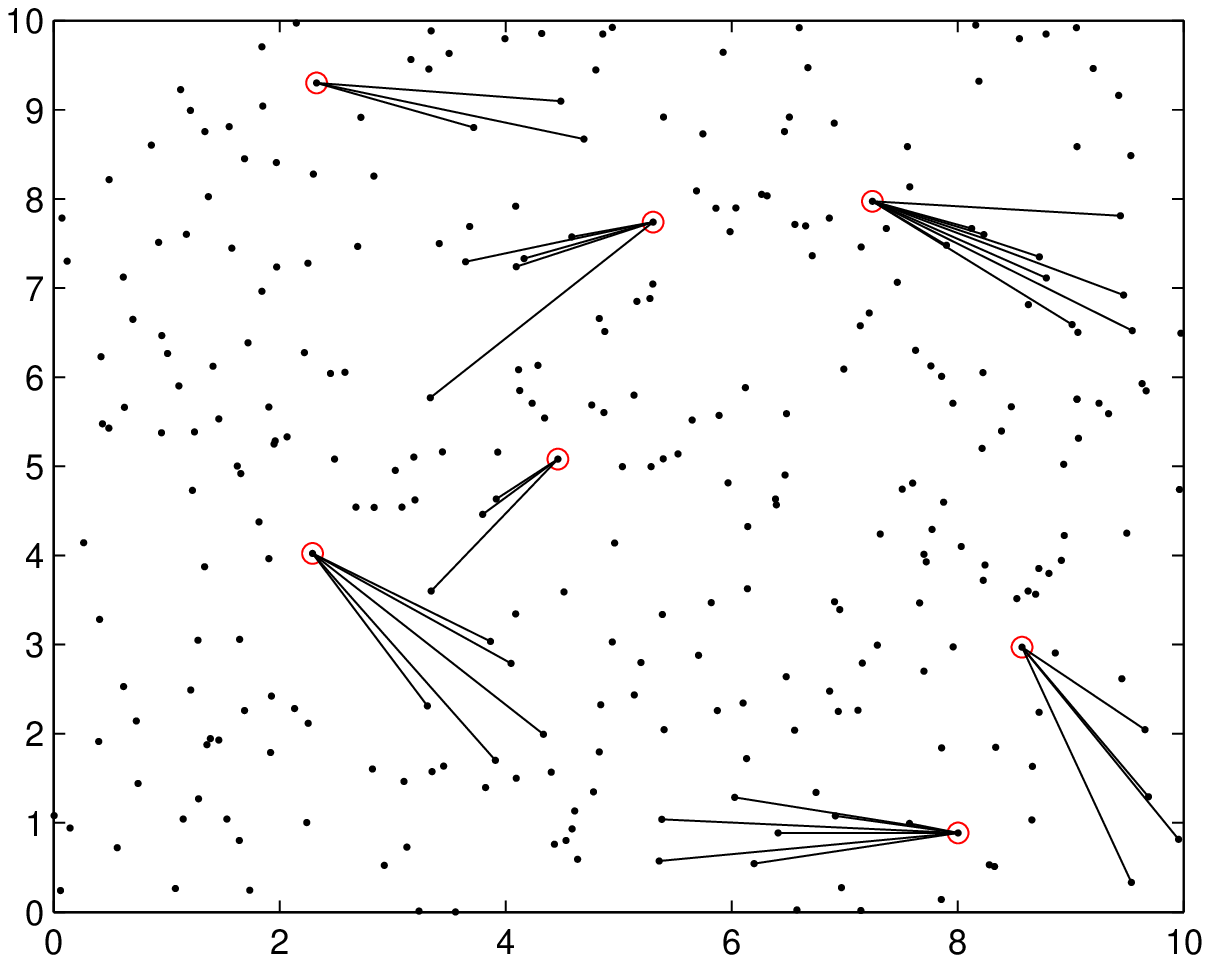}\label{subfig:nwillrnd_p001} }}
\mbox{\subfigure[Randomized Beams for $p=0.1$]{\includegraphics[width=3in]{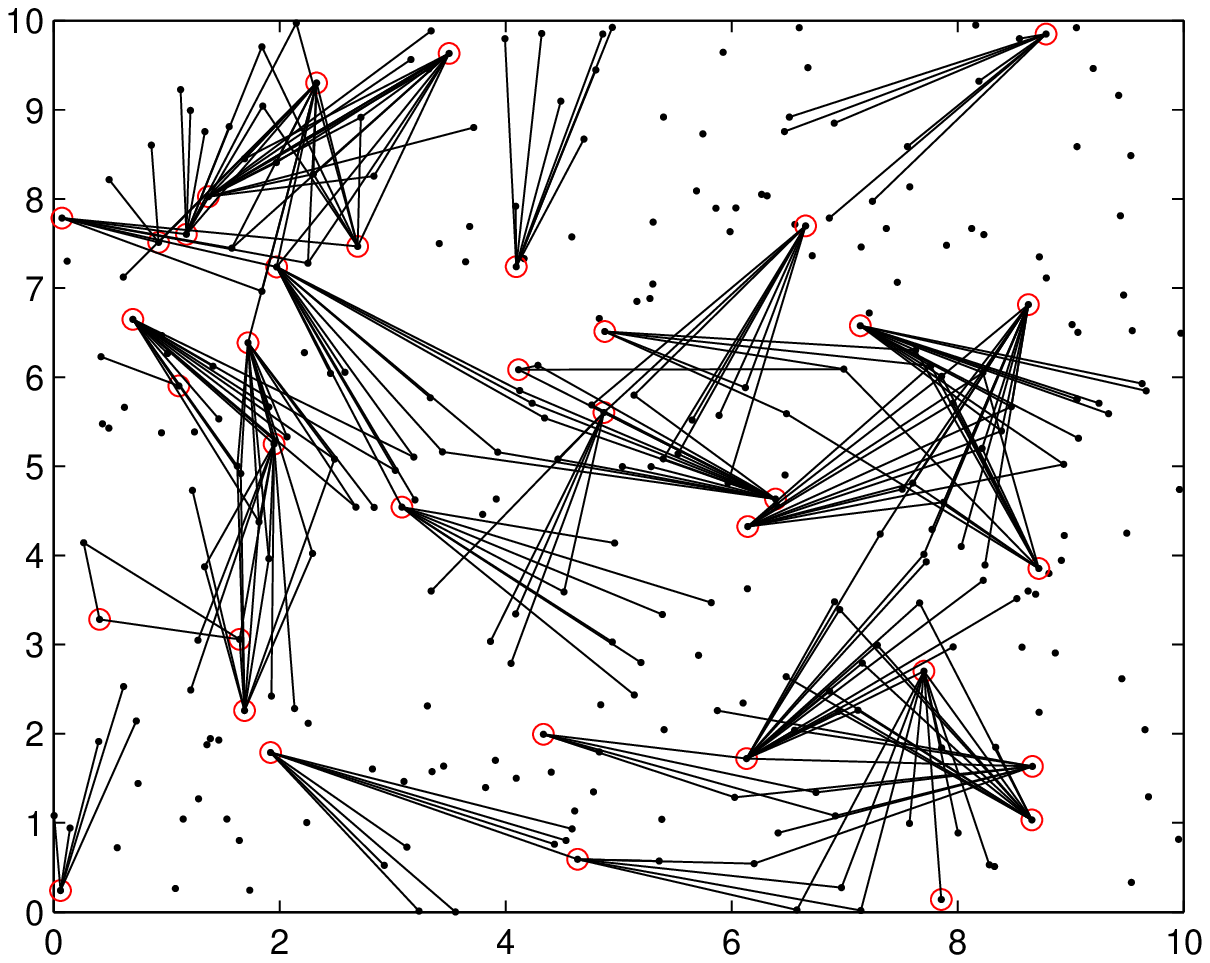}\label{subfig:nwillrnd_p01}}\quad
\subfigure[Randomized Beams for $p=1$]{\includegraphics[width=3in]{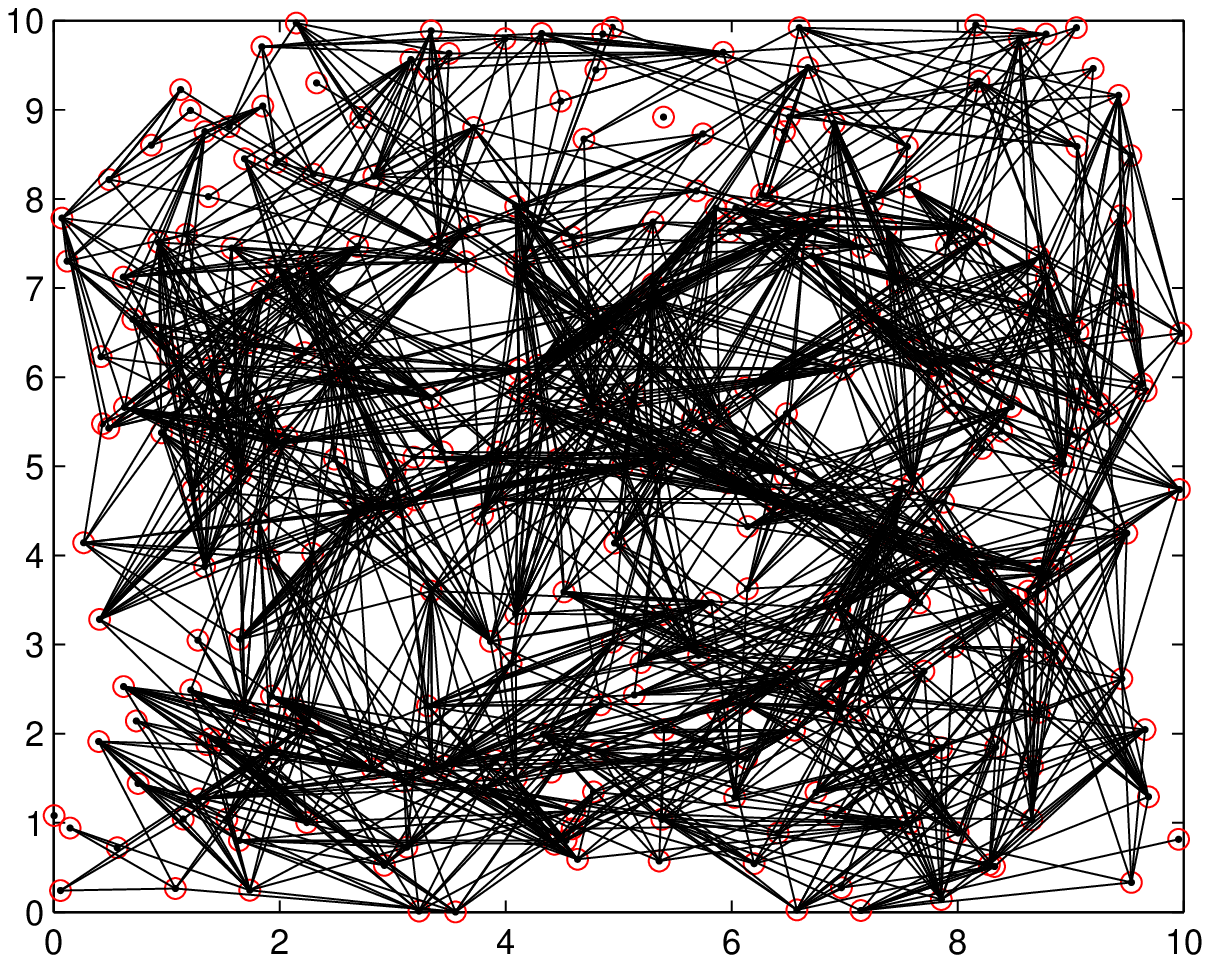}\label{subfig:nwillrnd_p1} }}
\caption{Illustration of network with randomized directional beams for different values of $p$.} \label{fig:nwillus_rndb}
\end{figure}

To account for the above tradeoffs, we choose a beamwidth that optimizes connectivity and beamlength depending on the density of nodes. We optimize over directional beam lengths that are integer multiples of the omnidirectional range $r$. To incorporate the tradeoff between increased length and connectivity, we divide the sector into separate regions of width $r$. Subsequently, we weigh the beam length $r(\theta)$ with the probability that at least one node is located in the first and the last regions,
\begin{equation}\label{eq:lC}
r_C = r(\theta) p_{nf} p_{nl}
\end{equation}
where $p_{nf}$ and $p_{nl}$ are the probabilities that at least one node is located in the first and last sectoral regions. The choice of the two probability terms is dictated by our motivation of maintaining connectivity while maximizing beamlength. Since an accurate estimate of maintaining bidirectional connectivity would require knowledge of the entire network, we use the term $p_{nf}$ to estimate the probability that connectivity is maintained with the omnidirectional neighborhood. As the first region under the beam lies within the omnidirectional range, a higher number of nodes here increases the probability that omnidirectional neighbors can be reached. Increasing the beamlength, however, is achieved by reduction in beamwidth. The narrow beamwidth reduces the probability of a node maintaining connectivity to its omnidirectional neighbourhood. The second probability $p_{nl}$ indicates the probability that at least one node benefits from the increased beam length. If there are no nodes present in the last region, greater connectivity can be achieved by increasing the beamwidth while the improvements from the beamlength stay the same. The nodes in the middle regions of the beam are not used in our expression since they do not represent the maximum benefits achievable by increased beamlength nor are they the most affected in terms of connectivity as a result of reduced beam width.

The optimum beam width $\theta^{*}$ among a set of values for $\theta$ is chosen as the one that maximizes $r_C$, i.e.
\begin{equation}\label{eq:lcondn}
\theta^{*} = \operatorname*{arg\,max}_{\theta} \left[r \sqrt{\frac{2 \pi}{\theta}}\right] p_{nf} p_{nl}
\end{equation}
The values for $p_{nf}$ and $p_{nl}$ are obtained based on the node density in the network. Given that the number of nodes in the omnidirectional neighbourhood of a node is $n$, the corresponding values are obtained as $p_{nf} = 1 - (1 - \frac{A_f}{\pi r^2})^n$ and $p_{nl} = 1 - (1 - \frac{A_l}{\pi r^2})^n$ where $A_f$ and $A_l$ are the area of the first and last regions respectively. Recall that the area under the beam is equal to that of the omnidirectional area when the same transmit power is used.

While the sector model is ideal for analysis, we also run simulations on a more realistic model of directional antennas to compare the performance. The model we use is that of a uniform linear array (ULA) \cite{BettstetterRndBeamform}, in which the antenna elements are arranged linearly. In a realistic scenario, the area under the beam depends on multiple factors including antenna characteristics and channel conditions. The beam pattern of a ULA is characterized by the number of elements used $m$ and the boresight direction $\theta_b$. An important difference with the sector model is that the beam pattern of ULA is characterized by one or more side lobes in addition to the main lobe. The maximum gain obtained in the direction of $\theta_b$ is equal to $m$. The signal transmitted by the node with a power $p_t$ is received correctly by a node located at a distance $s$ if the received power $p_r$ is greater than or equal to a power threshold $p_{r0}$. Given the transmitter and receiver gains $g_t$ and $g_r$ respectively, depending on the direction, $p_r$ depends on the power propagation environment, characterized by the pathloss exponent $\alpha$,
\begin{equation}\label{eq:pw-reln}
p_r = \frac{p_t g_t g_r}{s^\alpha}
\end{equation}
Beam patterns resulting from different values of $m$ and $\theta_b$ are illustrated in Fig. \ref{fig:beampatt_ula}. For a detailed discussion on antenna models, the reader is referred to existing literature on the subject in \cite{BettstetterRndBeamform,BalanisAntennaTh}. In order to map the sector model used earlier to ULA, we equate the number of antenna elements $m = \lceil \frac{r(\theta^{*})}{r} \rceil$, $r$ being the omnidirectional transmission range. We use $\alpha = 2$ while $g_r = 1$ as nodes always receive in omnidirectional mode. 
% where $r(\theta^{*})$ is the normalized beam length corresponding to $\theta^{*}$ obtained in equation (\ref{eq:lcondn}).

Fig. \ref{fig:nwillus_rndb} illustrates randomized beamforming using the sector model in a network with $300$ nodes for different values of $p$, which is the fraction of beamforming nodes.

\begin{figure}[tb]
    \centering
    \subfigure[Sector and ULA Models with Randomized Beamforming]{
    \includegraphics[scale=1.0]{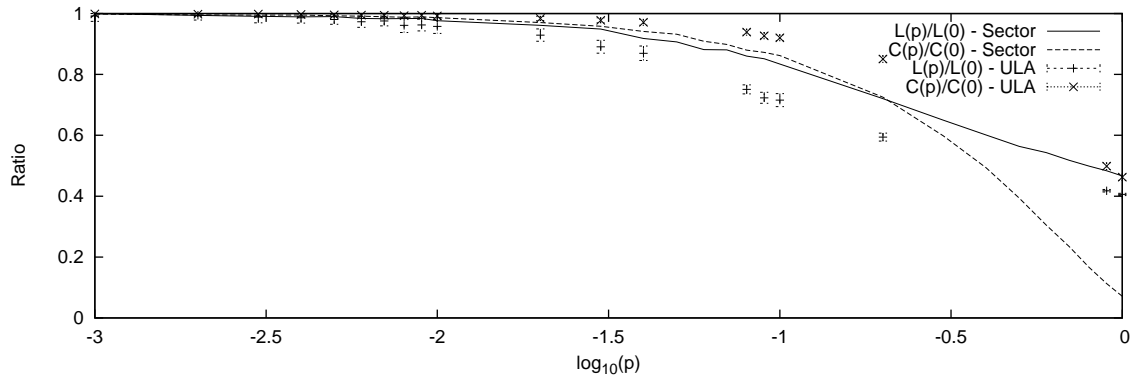}
    \label{subfig:aplcc_varpn300_rnd}
    }

    \subfigure[Comparing results of sector model with distance limited shortcut addition \cite{HelmySWWi}.]{
    \includegraphics[scale=1.0]{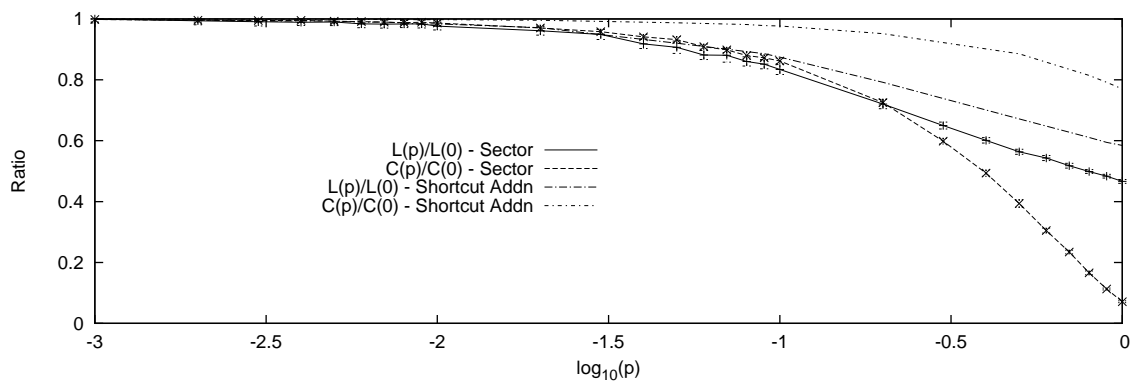}
    \label{subfig:aplcc_varpn300_sht}
    }

    \caption{Path Length Reduction and Clustering Coefficient as a function of varying probability of rewiring for $N=300$}
    \label{fig:varp_n300_aplcc}
\end{figure}

%\begin{figure}[tb]
%    \centering
%    \includegraphics[scale=1.0]{RatioCCPt_N300_secula}
%    \caption{Path Length Reduction and Clustering Coefficient as a function of varying probability of rewiring for $N=300$}
%    \label{fig:varp_n300_aplcc}
%\end{figure}

\begin{figure}
	\centering
    \includegraphics[scale=1.0]{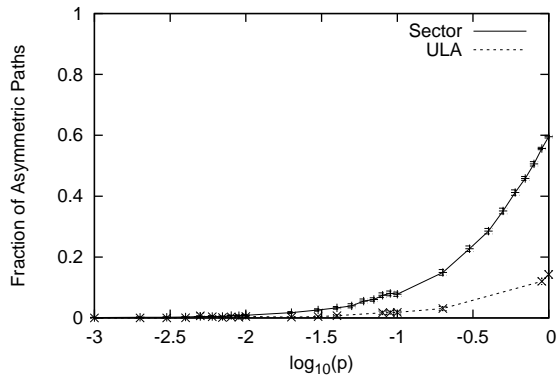}
    \caption{Asymmetric Paths as a function of varying probability of rewiring.}
    \label{fig:varp_n300_und}
\end{figure}

%\begin{figure}[tb]
%    \centering
%\mbox{\subfigure[Sector Model]{
%    \includegraphics[scale=1.0]{FracUniPairs_N300_log10}
%    \label{subfig:uni_varpn300}
%    }\quad
%    \subfigure[Uniform Linear Array (ULA) Model]{
%    \includegraphics[scale=1.0]{FracUniPairs_N300ULA_log10}
%    \label{subfig:uni_varpn300_ula}
%    }}
%
%    \caption{Asymmetric Paths as a function of varying probability of rewiring for $N=300$ using the ULA model.}
%    \label{fig:varp_n300_und}
%\end{figure}

%The results for the sector model, in Fig. \ref{subfig:aplcc_varpn300}, are compared to those using the ULA model, in \ref{subfig:aplcc_varpn300_ula}.
\subsection{Simulation Results}\label{subsec:rndbeam-sim}
For our simulations, we consider a network consisting of nodes using omnidirectional antennas distributed randomly in a rectangular region. We study the impact of using randomly oriented directional beams on the average path length (APL) and connectivity of this network. The simulations were run in MATLAB. The results shown were averaged over all possible node pairs in the network for $40$ different topologies.

For the first set of simulations, we vary the fraction of nodes $p$ that use directional beams while the rest of the nodes continue to use omnidirectional beams. The omnidirectional transmission range is normalized to $1$ with nodes distributed randomly in a $10$x$10$ region. Fig. \ref{subfig:aplcc_varpn300_rnd} shows the impact on path length reduction and the clustering coefficient for $N=300$. The corresponding effect on asymmetric paths is shown in Fig. \ref{fig:varp_n300_und}. The beam length $r(\theta^{*})$ obtained using equation (\ref{eq:lcondn}) results in a ratio $\frac{r(\theta^{*})}{D} \approx 0.2$. $D$ denotes the maximum distance between any two nodes in the network, i.e. the diameter of the network since we normalize $r$ to $1$. In Fig. \ref{subfig:aplcc_varpn300_rnd}, the ratio of the reduced path length $L(p)$ and clustering coefficient $C(p)$ to the initial values $L(0)$ and $C(0)$ are shown for both the sector and ULA models. We note that, for the sector model, the values for $\frac{L(p)}{L(0)}$ and $\frac{C(p)}{C(0)}$ are quite close to each other for low values of $p$. This is contrary to the desired results for small world networks as the path length reduction is accompanied by loss in connectivity. The adverse impact of $C(p)$ on connectivity is seen in Fig. \ref{fig:varp_n300_und} as a high percentage of nodes have asymmetric paths. However, when a more realistic ULA model is used, better results are obtained in terms of both the path length improvement and connectivity. The presence of side lobes implies that a beamforming node retains connectivity to a greater fraction of nodes in its omnidirectional neighbourhood, resulting in higher values of the clustering coefficient. This also accounts for shorter path lengths since, in contrast to the sector model, these nodes in the omnidirectional neighbourhood can now be reached in a single hop. As a result, it can be seen in Fig. \ref{subfig:aplcc_varpn300_rnd}, the reduction in path length is faster than that of the clustering coefficient. The relatively higher values of clustering coefficient result in better connectivity for lower values of $p$ when using the ULA model as shown in Fig. \ref{fig:varp_n300_und}. However, as the value of $p$ is increased, the problem of asymmetric paths is observed even in this case, though to a lesser extent than with the sector model.

To underscore the suitability of directional beamforming for small world creation, we view it in the context of existing literature. Helmy \cite{HelmySWWi} obtained results showing the impact of distance limited short cuts on the path length reduction. We compare these results for the corresponding value of $\frac{r(\theta^{*})}{D}$ to those of the sector model in \ref{subfig:aplcc_varpn300_sht}. We observe that a higher reduction in path length can be achieved using the sector model as compared to link addition though higher values of clustering coefficient are obtained for the latter. As shown earlier in Fig. \ref{subfig:aplcc_varpn300_rnd}, greater reduction in path length is achieved in the case of the realistic ULA model. 

\begin{figure}[tb]
    \centering
    \includegraphics[scale=1.0]{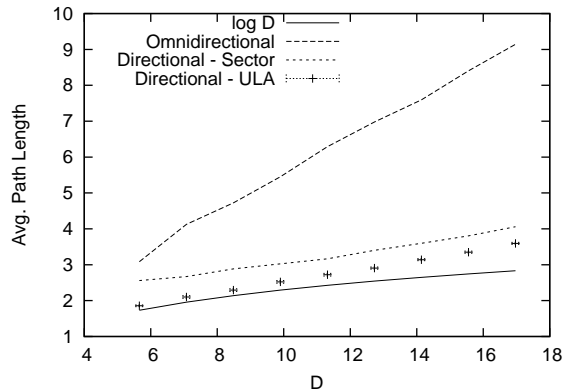}
    \caption{Growth of APL with increase in the size of the simulation region.}
    \label{fig:apl_vardiam}
\end{figure}

To observe the relation between the APL to the logarithm of the network size, we run a set of simulations in which the number of nodes is kept constant at $N=300$ while the size of the simulation region is increased. In Fig. \ref{fig:apl_vardiam}, the growth of the average path length when all nodes use directional beams is compared to the growth in the value of $\log D$ where $D$ is the largest possible distance between any two nodes in the network. The lines corresponding to the path length for both the sector and ULA models are parallel to that of $\log D$, indicating that the average path length grows as $O(\log D)$.

\subsection{Discussion and Insights}\label{subsec:insight}
An important feature of our results in the previous section is the tradeoff between path length improvement and connectivity in the network. As described earlier, our choice of beam length in equation (\ref{eq:lcondn}) is determined by maximizing the probability that a node maintains connectivity with its omnidirectional neighbours. However, as we see in the simulation results, connectivity is still affected severely when a higher fraction of nodes in the network use long range beams. 

% Objective; Relaxation of conn. limits wrt ULA; 
Based on the results from sector model alone, it is difficult to identify an optimum value of $p$ such that the network exhibits small world behaviour. Looking at the results using the sector model in Fig. \ref{subfig:aplcc_varpn300_rnd}, a $30 \%$ reduction in path length can be achieved with $p=0.2$ but results in  about $20 \%$ pairs being connected unidirectionally. Ensuring that almost all node pairs are connected bidirectionally would imply a value $p < 0.05$ but the corresponding improvement in path length is less than $10 \%$. However, looking at the results using the realistic ULA model, we observe that the constraints imposed by the sector model can be relaxed to an extent. Using the ULA can result in $30 \%$ reduction in the average path length with $p = 0.1$ while less than $2 \%$ of node pairs suffer from unidirectional connectivity. %This also provides us with additional insight for algorithm design. While the sector model is more tractable for algorithm design, we conclude that an algorithm that results in $20 \%$ of links being unidirectional is permissible as it is compensated in the realistic case.

While the simulations in this section considered randomly chosen nodes to beamform depending on the value of $p$, we would now like to investigate whether greater benefits can be achieved by deterministically choosing nodes to beamform. To do so, we seek to identify nodes in the network that are more likely to result in reduction of path length across the network if chosen to beamform. %We also use the insights obtained in this section to ensure that such a design does not result in loss of connectivity.
% In other words, we investigate whether the same fraction of nodes in the network can be used to give better results in terms of path length reduction and connectivity. 

\section{A Distributed Definition of Centrality for Wireless Ad Hoc Networks}\label{sec:dist-cent}
The structural importance of nodes in a network has traditionally been measured using different notions of centrality \cite{FreemanCentrality}. A commonly used measure of centrality is the betweenness \cite{FreemanBw} which measures the occurrence of a node along paths between other node pairs in the network. High betweenness of a node indicates that it lies along a majority of paths in the network. Creating shortcuts on nodes with high betweenness values is likely to maximize reduction in the average path length in the network as a majority of paths are affected. Various definitions of betweenness centrality have been proposed in the existing literature. Betweenness centrality, as defined in \cite{FreemanBw}, is known as the Shortest Path Betweenness Centrality (SPBC) as it estimates the importance of a node with respect to the shortest paths between all other nodes in the network. Variants of the SPBC were reviewed by Brandes in \cite{BrandesSPBC}. Everett and Borgatti \cite{EverettEgoBw} proposed the Ego Betweenness Centrality which is calculated individually by nodes using only their immediate neighborhood information. An alternate class of betweenness centralities that do not consider the shortest paths in the network was proposed by Freeman et al. in \cite{FreemanFBC}. Flow Betweenness Centrality (FBC) measures the importance of a node based on traffic flow in the network. Routing Betweenness Centrality (RBC), proposed by Dolev et al. in \cite{DolevRBC}, generalizes SPBC and FBC by considering paths resulting from routing strategies. % Improve

For our problem of distributed self-organization of the network, we need nodes to accurately estimate their importance with respect to paths in the network. This constraint makes it unsuitable to use any of the conventional measures as they consider a centralized view of the network. In addition to the distributed computation of the centrality itself, nodes need to decide on their beamforming behavior based on these values. As nodes with high betweenness are ideal for shortcut creation, nodes need to estimate the rank of their betweenness in the network. In order to satisfy these requirements, our algorithm design is centered on a new measure of centrality, which we define in this paper, that allows nodes to decide on shortcut creation depending on the structure of the network. 

\subsection{Wireless Flow Betweenness}\label{subsec:wfb-defn}
The computation of betweenness in a network has been shown to scale as a function of the longest geodesic \cite{RavaszScaling}. In wireless ad hoc networks, the computation costs are particularly costly because of the transmission overheads involved. In order to ensure that network performance is not affected, thus, the computation of a betweenness measure should ideally incur zero overheads. Using this as the motivation, we propose a measure of betweenness computed by nodes by exploiting the wireless broadcast advantage (WBA), thereby minimizing the overheads incurred.

The proposed measure of betweenness, Wireless Flow Betweenness (WFB), makes use of information extracted from traffic flows being routed through the network to estimate the structural importance of nodes in the network. This allows the computed measure of centrality to adapt to the traffic flow patterns in the network. Further, traffic flow information can be obtained locally by nodes, thereby minimizing the transmission overheads. Finally, such a mechanism can proceed in parallel with regular network operations such as routing and packet delivery, therefore increasing its robustness. We structure our discussion of defining WFB in a manner that makes the reasoning behind the proposed measure clear to the reader. We start off with a simplistic expression of locally computable centrality. Subsequently, we identify the redundancies involved in such a measure and describe how we address them. As part of this, we also refer to WFB measure we had proposed earlier in \cite{BanerjeeWFBCCNet} and how the definition in this paper addresses the deficiencies in \cite{BanerjeeWFBCCNet}.

The broadcast nature of the wireless medium results in implicit sharing of information as nodes can overhear all transmissions in their one-hop neighborhood. Based on this neighborhood information alone, a node can estimate its importance in terms of how often it transmits itself, either as a source or as a forwarding node. A straightforward measure of betweenness of a node $v$, can therefore be computed as the ratio of the number of times it acts as a forwarding node to the total number of unique traffic flows in its neighborhood. This can be expressed as,
\begin{equation}\label{eq:bw-simp}
w(v) = \frac{g(v)}{\sum\limits_{u \in \{\mathcal{N}(v) \cup v\}} g(u)}
\end{equation}
where $g(u)$ denotes the number of packets forwarded by a node $u$ for distinct source-destination pairs while $\mathcal{N}(v)$ denotes the set of neighbours of $v$. The denominator gives the total number of packets forwarded in the neighbourhood of $v$. % Each node records the number of packets forwarded by each of its neighbours.

The expression in (\ref{eq:bw-simp}), though straightforward in design, only gives an egocentric measure of importance of the node $v$. This is because a count of the transmissions in the neighborhood does not give any indication of the likelihood of any flow in the network passing through $v$. Identifying the exact number of traffic flows in the entire network, including those that do not pass through the neighborhood of $v$, would be non-trivial and costly. Instead, we propose estimating the betweenness of a node by propagating network information as part of traffic flows. We propose the Wireless Flow Betweenness (WFB) which uses recursive computation at nodes to obtain an accurate estimate.

The betweenness of a node $v$ based on the set of traffic flows in the network, was defined in \cite{FreemanFBC} as the Flow Betweenness Centrality (FBC) which refers to the share of the maximum flow between all node pairs in the network that passes through $v$. However, as mentioned earlier, we cannot adapt FBC directly as doing so would require nodes to maintain global network state. A recently proposed measure of centrality for wireless ad hoc networks \cite{MaglarasPathBased} requires knowledge of multihop neighborhood of a node, which is again costly in terms of transmission overheads. In order to minimize the state information, our approach centers around estimation of the number of flows by each node using neighbourhood values of centrality.

In \cite{BanerjeeWFBCCNet}, we proposed a WFB measure in which estimation of the number of flows is done by simply inverting the expression in (\ref{eq:bw-simp}). For any node $u$, the number of transmissions it is aware of is obtained as $\frac{g(u)}{w(u)}$. Thus, a node $v$ can gain an estimate of the number of transmissions in the network using the above estimation for each of its neighbors $u \in \mathcal{N}(v)$. Nodes piggyback their self-computed values of centrality whenever they transmit a packet, either as a forwarding node or as source. However, this involves multiple redundancies that can result in inaccurate estimation, which we discuss here. % In the WFB measure proposed earlier in \cite{BanerjeeWFBCCNet}, the centrality values of nodes were obtained using the sum of the estimated number of transmissions for each neighbour.In the rest of the discussion, we use $w(u)$ to denote the betweenness of a node $u$ so as to distinguish it from the measure obtained in (\ref{eq:bw-simp}).
% First redundancy: $v$ is aware of a subset of flows that $u$ is aware of, i.e. those that are forwarded by $u$
% Second redundancy: Multiple nodes in $N(v)$ are also aware of each other's transmissions and affect each other's centrality values
% fact that the node $v$ is itself aware of a subset of traffic flows it estimates for a neighbor $u$ with the expression $\frac{g(u)}{w(u)}$. This is because $v$ can overhear all packets transmitted by $u$, $g(u)$. 

The first redundancy arises due to the term $g(u)$ that gives the number of packets transmitted by a node $u$, which lies in the neighbourhood of $v$. As $v$ can overhear all of $u$ transmissions, it counts all of these as part of the number of traffic flows $g(v)$ it counts in its neighbourhood and uses to compute $w(v)$. Thus, $g(v) \geq g(u) \forall u \in \mathcal{N}(v)$. As $g(u)$ is counted again as part of the expression $\frac{g(u)}{w(u)}$, this results in overestimation. Thus, node $v$ needs to ensure that the term $g(u)$ is not counted multiple times so as to accurately estimate the number of traffic flows which $u$ is aware of but $v$ is not. Therefore, we estimate the number of additional flows that $u$ is aware of as
\begin{equation}\label{eq:af}
a_f(u) = \frac{g(u)}{w(u)} - g(u)
\end{equation}
The betweenness of $v$ can be computed using the $a_f$ values of all its neighbors as,
\begin{equation}
w(v) = \frac{g(v)}{o(v) + \sum\limits_{u \in \mathcal{N}(v)} a_f(u)} \nonumber
\end{equation}
where $o(v)$ denotes the number of traffic flows overheard by $v$, including those transmitted by $v$ itself. A node updates its betweenness if it either transmits a packet or overhears one from its neighbors. 

However, the above summation over all neighbors of $v$ introduces a second redundancy which needs to be taken care of. Just as $v$ is aware of all packets transmitted by a neighbor $u$, multiple neighbors of $v$ also affect each other's betweenness values. Neighbors of $v$ which are also neighbors of each other estimate the term $a_f$ based on the values of $w$ observed for each other. Consider two nodes $u$ and $u'$ which are neighbors of $v$ as well as neighbors of each other. As part of their centrality calculations, $u$ and $u'$ estimate $a_f(u')$ and $a_f(u)$ respectively. Both these values include the number of transmissions made by $v$. In the absence of a one-hop neighborhood map, neither node is aware of $v$ being a common neighbor and therefore, $a_f(v)$ is implicitly counted more than once, resulting in overestimation of the number of flows at all the three nodes. This effect is aggravated over multiple iterations and for networks with high node density.

In order to handle the second redundancy, we limit the set of neighbors whose betweenness values are used for computation. Nodes that infrequently act as forwarding nodes spend the majority of time overhearing transmission from other nodes. Computing the $a_f$ measure for such a flow is likely to result in a lot of redundancy as the majority of estimated additional flows are likely to be redundant. Thus, we revise the definition of Wireless Flow Betweenness (WFB) to only consider the neighbor which acts as a forwarding node most often. The possibility of redundancy is reduced with such a choice as the node chosen is unique within a neighborhood. Further, since a majority of traffic in a given network region flows through this node, it gives a better estimate of the number of flows. Based on this discussion, the WFB value of a node $v$ is obtained as,
\begin{align}\label{eq:wfb-final}
w(v) =& \frac{g(v)}{o(v) + (\frac{g(u)}{w(u)} - g(u))} \\
&\mbox{where } u = \operatorname*{arg\,max}_{u' \in \mathcal{N}(v)} g(u') \nonumber
\end{align}
A node computes its WFB value whenever it acts as a forwarding node or overhears a transmission in its neighborhood. 

By considering a single node in the neighborhood, we reduce the chances of overestimation. However, it also raises the possibility of some flows not being counted. Looking at the denominator of the expression in (\ref{eq:wfb-final}), we observe, as before, that the term $o(v)$ accounts for all transmissions in the neighborhood of $v$, and therefore includes $g(u)$. Thus, the term $(\frac{g(u)}{w(u)} - g(u))$ gives the number of additional flows in the network estimated recursively. As with node $v$, the value of $w(u)$ takes into consideration the number of flows overheard by $u$ as well as the node with maximum forwarding count in $\mathcal{N}(u)$. The consideration of the neighbor with maximum forwarding count implies that a majority of flows are always accounted for in a neighborhood, thereby minimizing the probability of flows not being counted. We note that this estimation over successive hops can stop when a node itself has the maximum forwarding count in its neighborhood, leading to an estimation error. On further consideration, however, we observe that this is unlikely to occur unless the node is the global maxima, i.e. has the maximum centrality value in the entire network. Thus, the possibility of inaccuracies in estimation is minimized. We verify this in the next section based on the rank correlation of WFB values in the network with that of centrally computed FBC.

\begin{figure}[tb]
    \centering
    \includegraphics[scale=1.0]{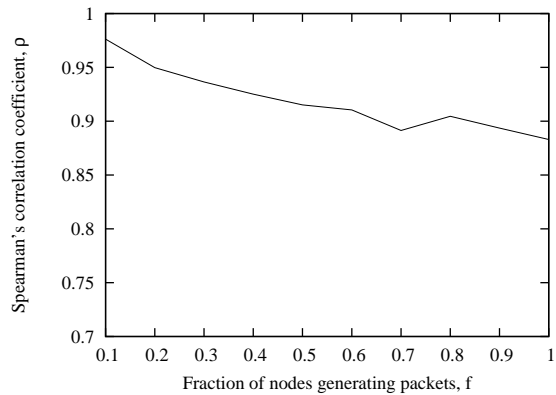}
    \caption{Correlation between WFB and FBC for varying traffic load.}
    \label{fig:rho_fbc}
\end{figure}

\subsection{Correlation with Flow Betweenness Centrality (FBC)}\label{subsec:comp-wfbfbc}
To verify the validity of the proposed WFB measure, we compare the values obtained with the Flow Betweenness Centrality (FBC) \cite{FreemanBw}. Since the relative importance of a node with respect to either betweenness measure is given by its rank in the network, we obtain the rank correlation of the two measures. We rank the nodes in the network separately according to their WFB and FBC values. Subsequently, we measure the correlation using the Spearman's correlation coefficient,
\begin{equation}\label{eq:spear-rho}
\rho = 1 - \frac{6 \sum\limits_{i \in V} d_i^2}{n(n^2 - 1)}
\end{equation}
where $V$ is the set of all nodes in the network and $d_i$ is the absolute difference in rank between the two rankings for the $i$-th node. A value $\rho = 1$ implies perfect correlation between the two rankings.

We vary amount of traffic in the network and obtain the value of $\rho$ for the WFB and FBC values obtained thereafter. A fraction $f$ of all nodes in the network generate a packet to randomly chosen node as the destination. The results are shown in Fig. \ref{fig:rho_fbc}. A very high correlation is seen for lower values of $f$ but it drops slightly for higher values. Since the computed WFB measure does not make use of explicit transmissions to propagate information across the network, as $f$ increases to $1$, a greater percentage of information does not propagate over multiple hops. This results in lower values of $\rho$ as $f$ increases. However, even for high values of $f$, $\rho > 0.88$ implies high correlation between WFB and FBC.

\subsection{Overhead and Buffer Costs}\label{subsec:ovhdbuf-cost}
While the above results show that there is a close correlation between the WFB values computed at individual nodes and the corresponding FBC values, it is necessary to understand the additional costs required for such computation. 

The transmission overhead costs are minimal since only WFB values are piggybacked on to packets by the forwarding nodes, involving one additional field. Thus, additional transmission costs are of constant order. Our design, however, requires nodes to store the WFB values of their neighbors along with the corresponding forwarding count. The buffer requirements, thus, scale with increase in the node density. Given a neighborhood size of $n$ nodes, a node needs to store three fields for each neighbor, namely the node identity along with the WFB value and the forwarding count, resulting in a buffer size of the order of $O(3n)$. However, since a lot of existing ad hoc network mechanisms rely on the presence of neighborhood knowledge, which is of $O(n)$, the additional costs involved for computation of WFB are unlikely to impose a significant burden.

% Talk about buffer costs for storing flow information? - Check the implementation again

\begin{figure}
\centering
\mbox{\subfigure[Initial Network Setup with Omnidirectional Beams]{\includegraphics[width=3in]{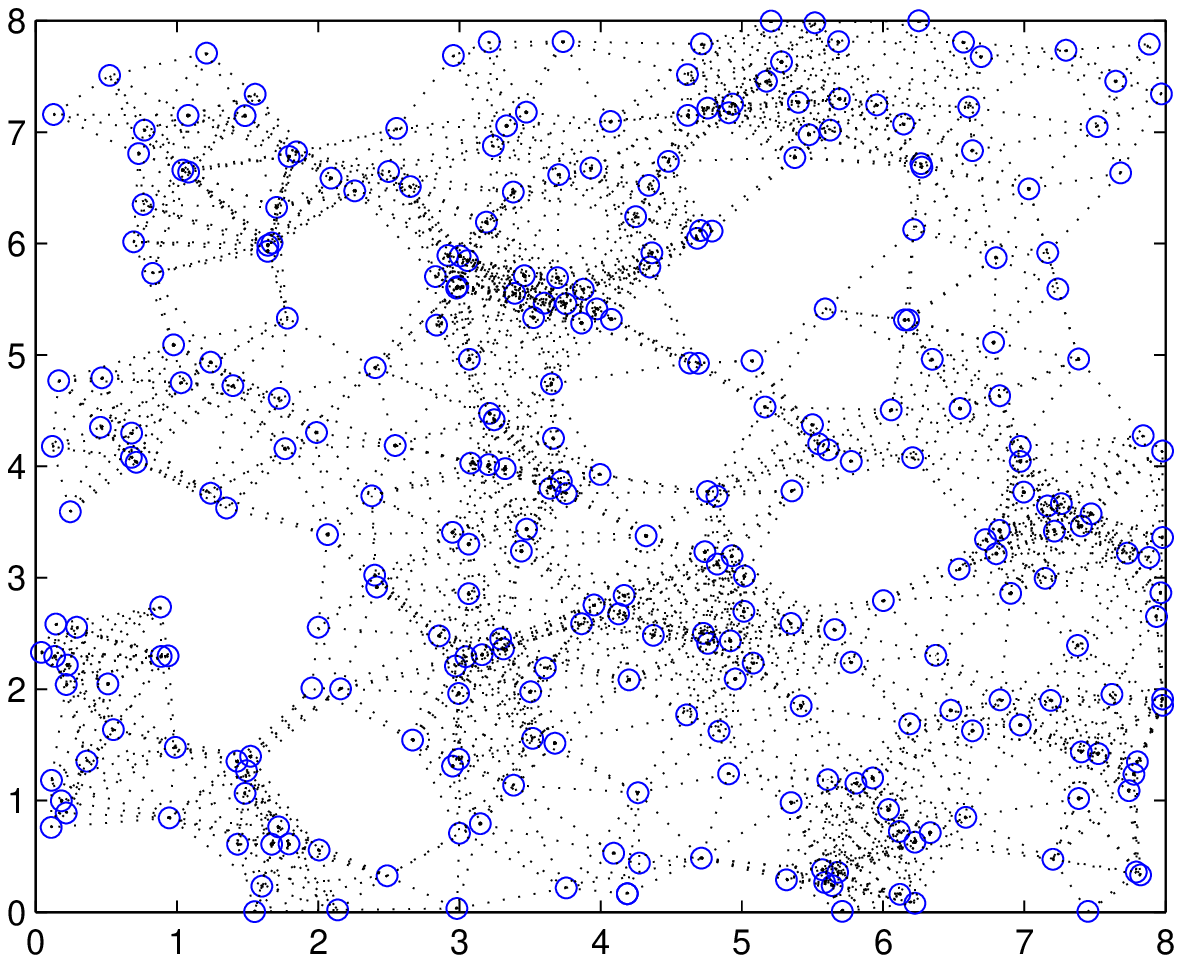}\label{subfig:nwill_initnw}}\quad
\subfigure[Centralized choice of nodes ranked by WFB values]{\includegraphics[width=3in]{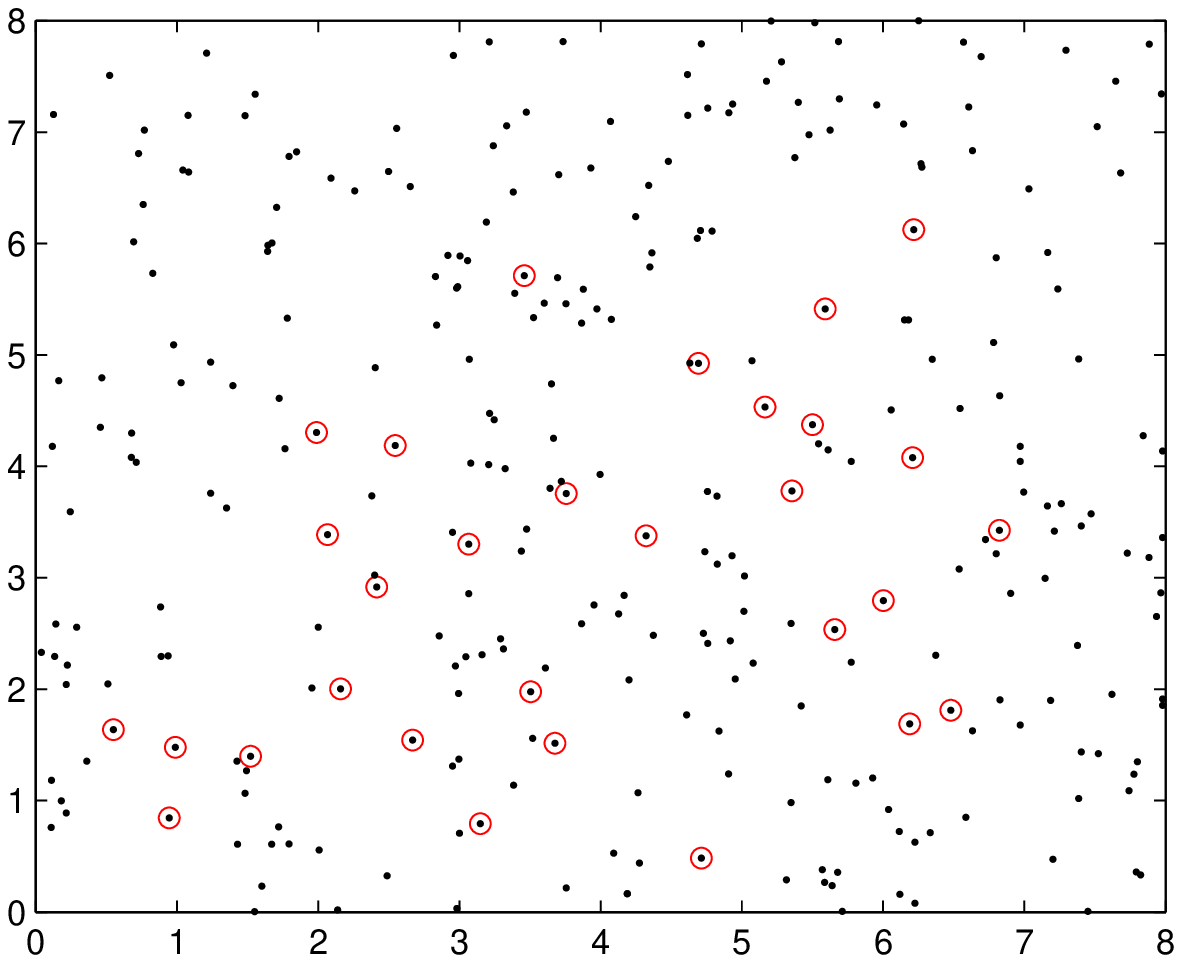}\label{subfig:nwill_ndchctr} }}
\mbox{\subfigure[Edges with sector model for centralized choice of nodes]{\includegraphics[width=3in]{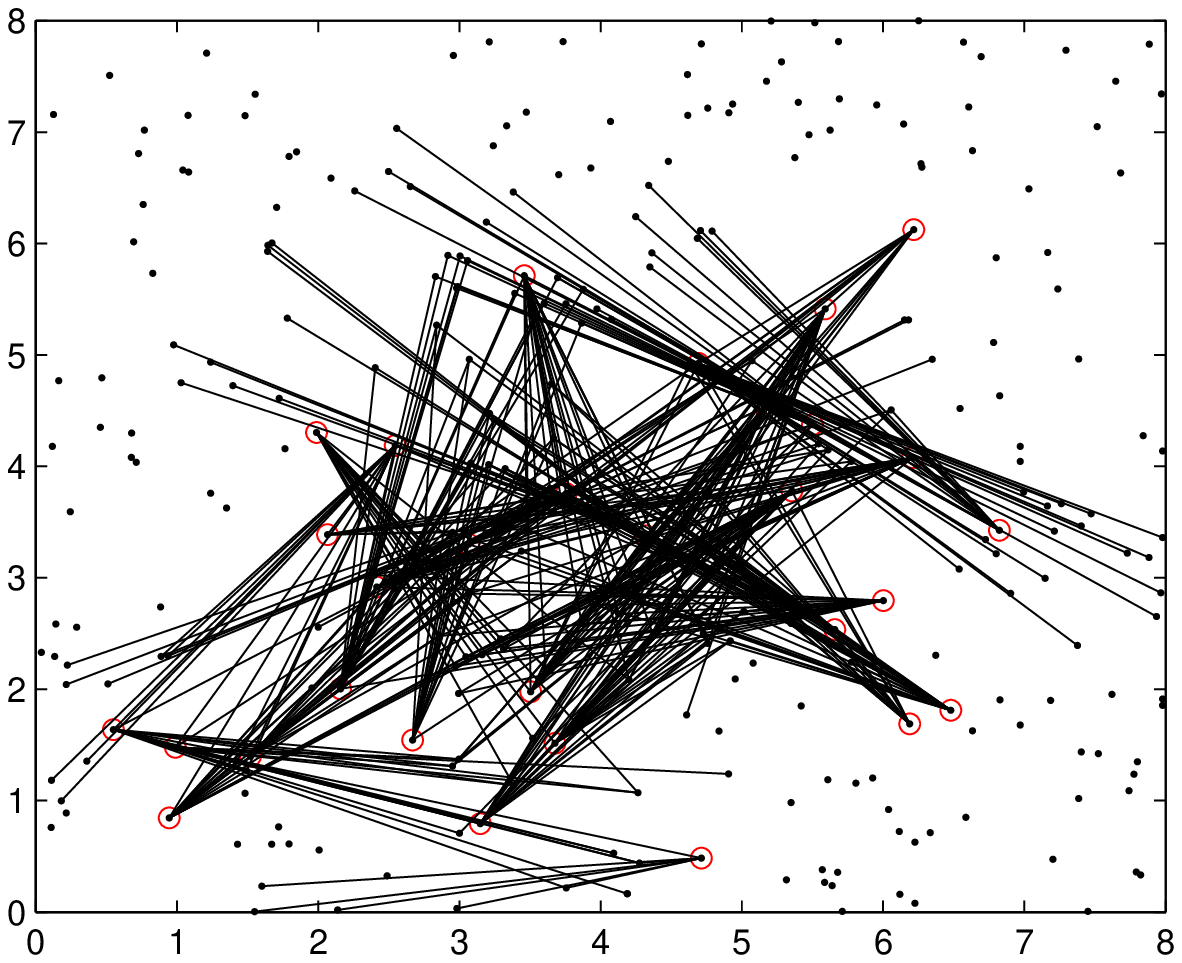}\label{subfig:nwill_ndchctrbms}}\quad
\subfigure[Distributed choice of beamforming nodes]{\includegraphics[width=3in]{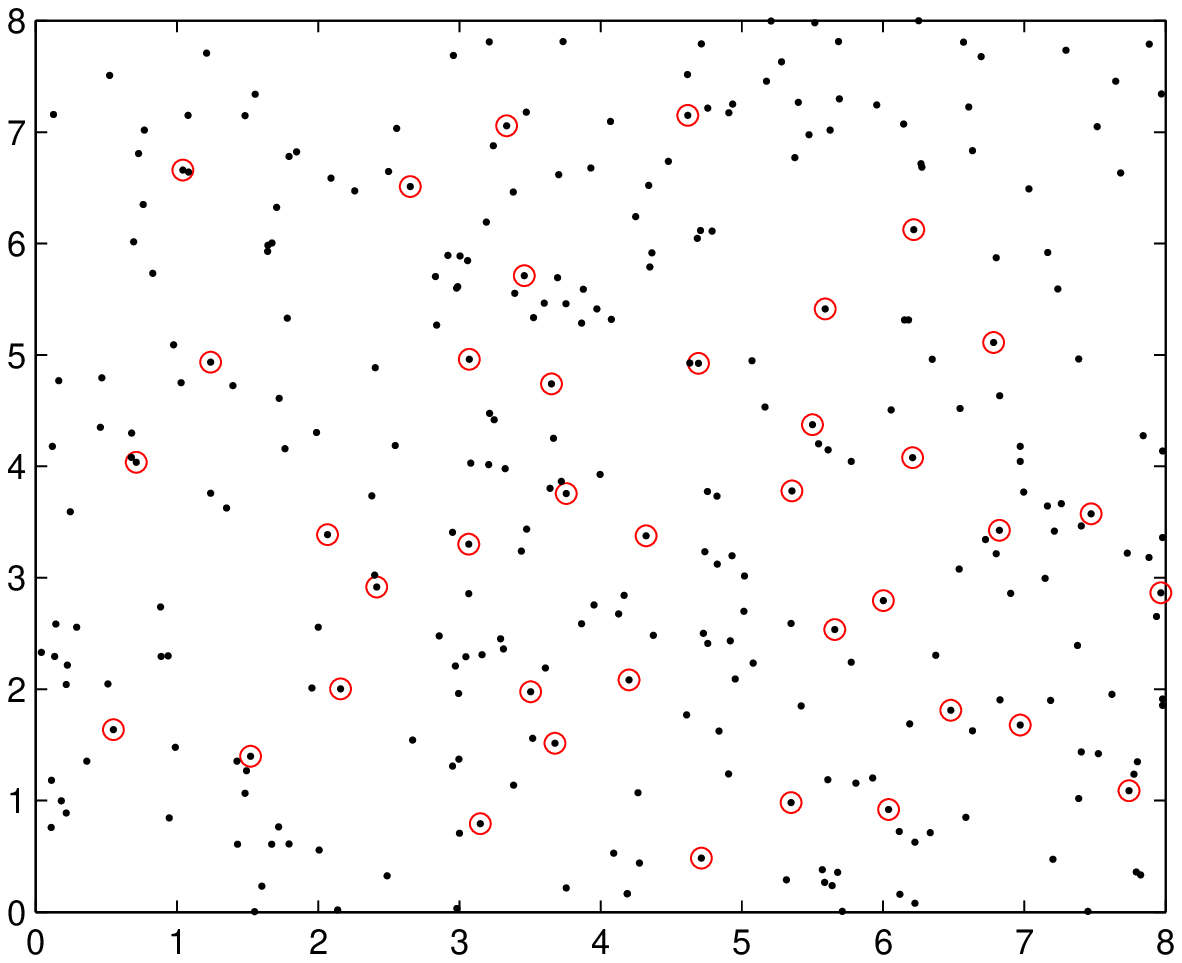}\label{subfig:nwill_ndchdst} }}
\mbox{\subfigure[Edges corresponding to sector model for distributed choice of nodes]{\includegraphics[width=3in]{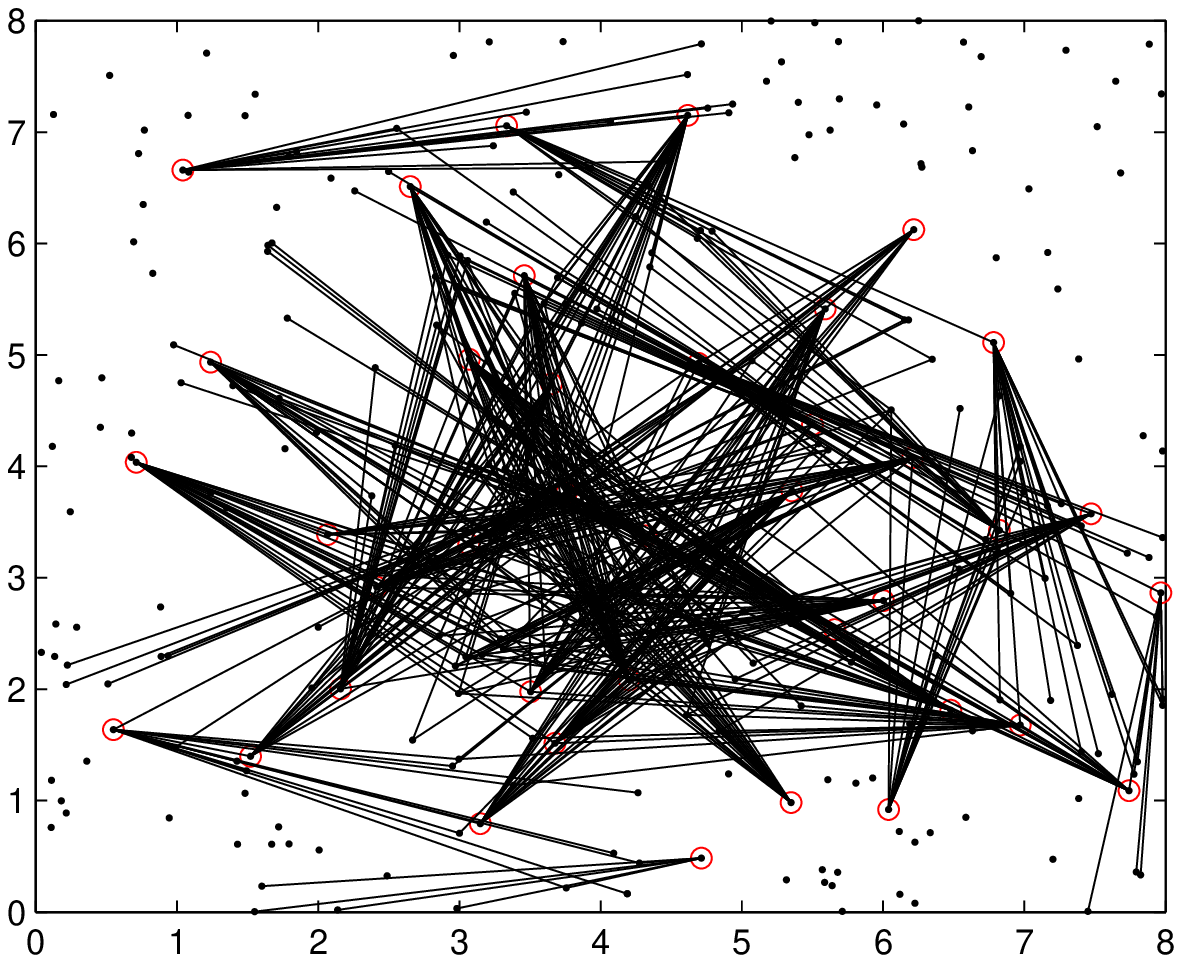}\label{subfig:nwill_ndchdstbms}}\quad
\subfigure[Edges corresponding to ULA model]{\includegraphics[width=3in]{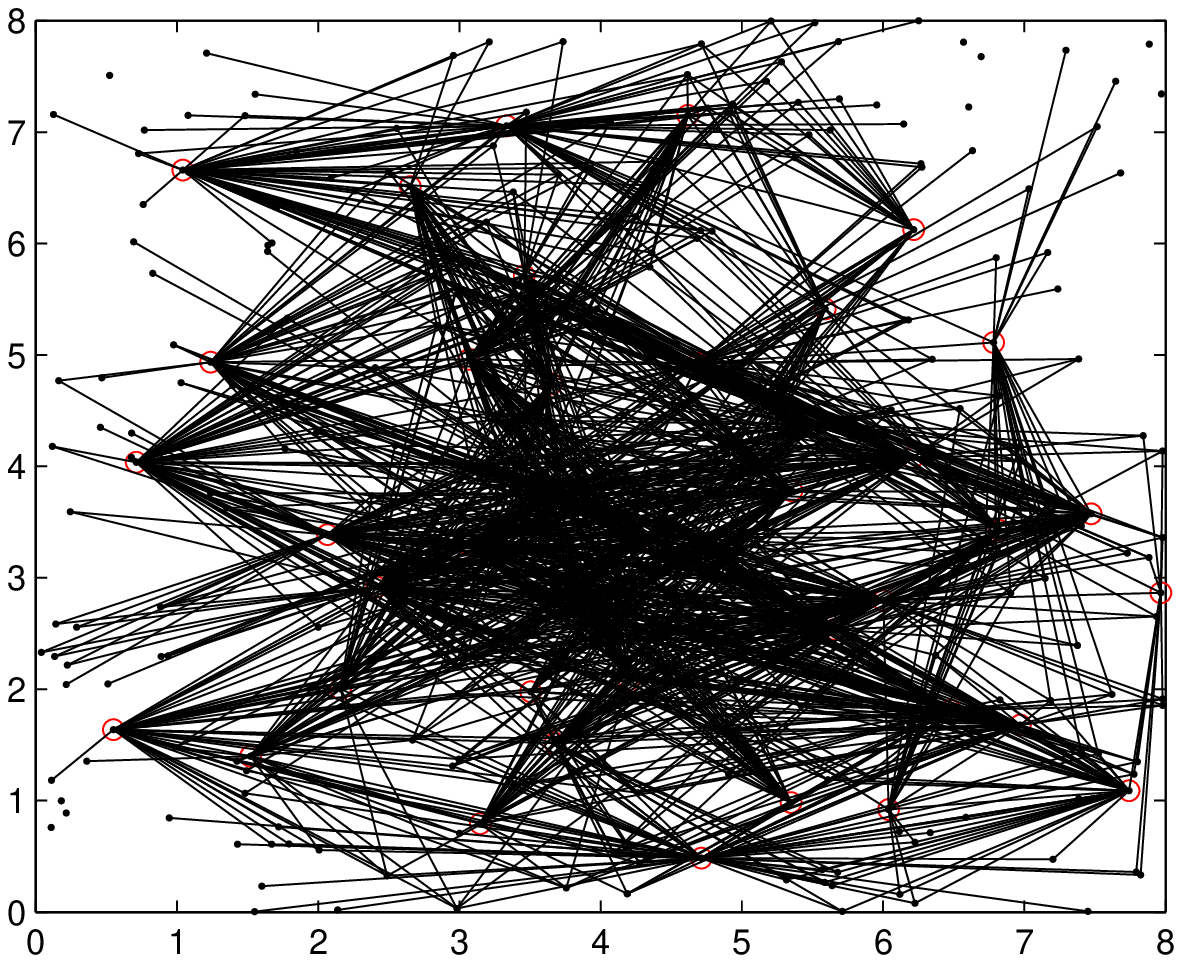}\label{subfig:nwill_ndchdstula} }}
\caption{Illustration of reorganization of network using centralized and distributed choices of beamforming nodes using WFB.} \label{fig:nwillus_wfb}
\end{figure}

\subsection{Using WFB for Small World Creation with Beamforming}\label{subsec:wfb-swn}
We now explore the use of WFB for small world creation using directional beamforming. Motivated by the close correlation between WFB and FBC, we evaluate the performance benefits of using nodes with high values of WFB as beamforming nodes. Note that here we do not discuss the issue of distributed identification of beamforming nodes. Instead, we consider that somehow the top ranking nodes with respect to WFB values are identified and create beams. Based on the insights obtained in this section, we formulate an algorithm for distributed beamforming in the next section.

\begin{figure}[tb]
    \centering
\mbox{\subfigure[Path Length Reduction]{
    \includegraphics[scale=1.0]{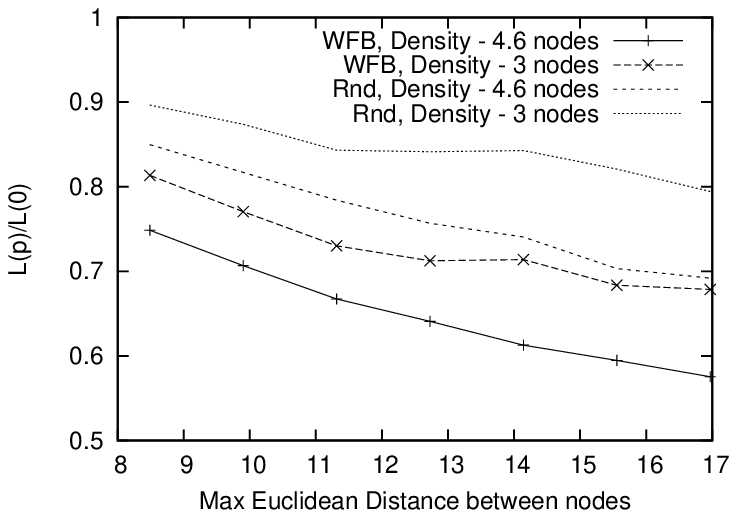}
    \label{subfig:ptlrat_wfbsw_compdns}
    }\quad
    \subfigure[Growth of average path length]{
    \includegraphics[scale=1.0]{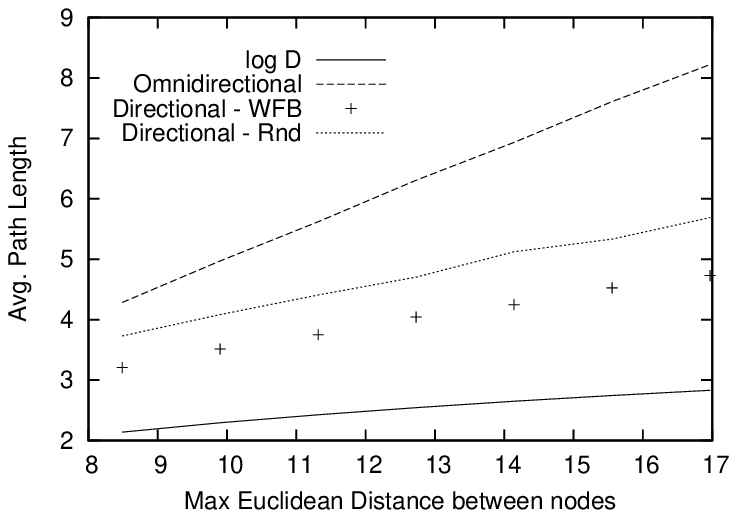}
    \label{subfig:ptgrwth_wfbsw}
    }}

    \caption{Path length reduction by beamforming at nodes with high values of WFB.}
    \label{fig:vardns_wfbsw_ptl}
\end{figure}

We consider that the top $10 \%$ (i.e. $p = 0.1$) of all nodes in the network create directional beams. The beamwidth used is the same as for randomized beamforming, formulated in equation (\ref{eq:lcondn}). Nodes orientate their beams in a direction in which they record the maximum hop count based on earlier traffic flows, so as to minimize the network diameter. The simulation results shown here only consider the sector model. We compare results for nodes distributed at two different node densities. A high node density of $d = 4.6$ nodes per unit area results from distributing $300$ nodes over an 8x8 region, and a lower density of $d = 3$ is obtained by increasing the region size to 10x10 for the same number of nodes. Subsequently, we keep the density constant at either of these two values and vary the size of the region.

We show the performance improvements available for increasing size of the network region, indicated using $D$ as the maximum distance between any two nodes. Fig. \ref{subfig:nwill_ndchctr} shows the set of nodes chosen as beamforming nodes for the network setup shown in \ref{subfig:nwill_initnw}. It can be seen that the majority of the nodes chosen (circled in red) are located towards the centre of the network. Edges corresponding to the directional beams oriented as specified above are shown in Fig. \ref{subfig:nwill_ndchctrbms}.

\begin{figure}[tb]
    \centering
    \includegraphics[scale=1.0]{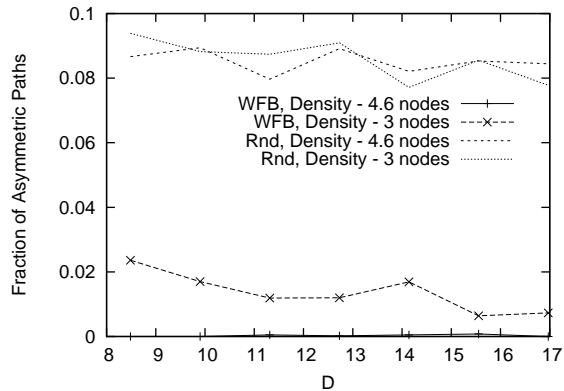}
    \caption{Impact on unidirectional connectivity.}
    \label{fig:vardns_wfbsw_und}
\end{figure}

Fig. \ref{fig:vardns_wfbsw_ptl} illustrates the impact on path length reduction. A reduction of more than $40 \%$ and $30 \%$ are shown to be achievable for $d = 4.6$ and $d = 3$ respectively in Fig. \ref{subfig:ptlrat_wfbsw_compdns}. Here, it is notable that the improvements in both cases are higher than the corresponding value of $p$ for randomized beamforming, as shown in section \ref{sec:swnda-anal}. As the network size increases, the beam length to diameter ratio reduces to $\frac{r(\theta^{*})}{D} \approx 0.2$ for $d = 3$ and $\frac{r(\theta^{*})}{D} \approx 0.25$ for $d = 4.6$. Using the results in \cite{HelmySWWi} as a benchmark, we see that the path length reduction is greater for corresponding values of $\frac{r}{D}$. The results in Fig. \ref{subfig:ptgrwth_wfbsw} compare the growth of the average path length with the logarithm of the network size. The line corresponding to reduced path length using directional beams is shown to grow in parallel with the line corresponding to $\log D$, implying that the growth in path length is logarithmic to that of the network size.

Fig. \ref{fig:vardns_wfbsw_und} shows the impact of directional beamforming using WFB on the fraction of node pairs that are unidirectionally connected. The effect on unidirectional connectivity is negligible for $d = 4.6$. For $d = 3$, a relatively higher fraction of node pairs are connected unidirectionally, though this is still lower than the value for randomized beamforming. The improvement in connectivity results from the fact that nodes with high value of centrality are better connected. For lower node density, the higher fraction of unidirectional connectivity results due to our choice of beamforming nodes. Nodes with high values of WFB are likely to be neighbors of each other as they are located towards the centre of the network. As we choose nodes with high values of WFB as beamforming nodes, they are likely to be neighbors to each other. In the case of lower node densities, this results in a higher fraction of neighboring nodes which are beamforming thereby resulting in unidirectional connectivity. This is alleviated for higher node densities as, in spite of neighboring nodes creating beams, a higher number of neighbors continues to use omnidirectional beams thereby increasing the chances of bidirectional paths.

\section{Distributed Small World Creation using Wireless Flow Betweenness (WFB)}\label{sec:swn-wfb-algo}
We now focus on distributed algorithm design for small world creation using directional beamforming. Our design centers around nodes determining their beamforming behavior based on their estimated importance in the network, thereby adapting small world creation to the network structure. We use Wireless Flow Betweenness (WFB), defined above, to identify the optimal set of nodes to beamform for achieving small world behavior.

In the previous section, we illustrated the benefits achievable by using top ranked nodes by WFB as beamforming nodes. However, as the set of nodes with highest values of WFB was identified using global network information, it is not suitable for distributed implementation. Further, as discussed before, only choosing nodes with high values of WFB can result in unidirectional connectivity at low node densities. 

\begin{figure}[tb]
    \centering
\mbox{\subfigure[Distribution of WFB values]{
    \includegraphics[scale=1.0]{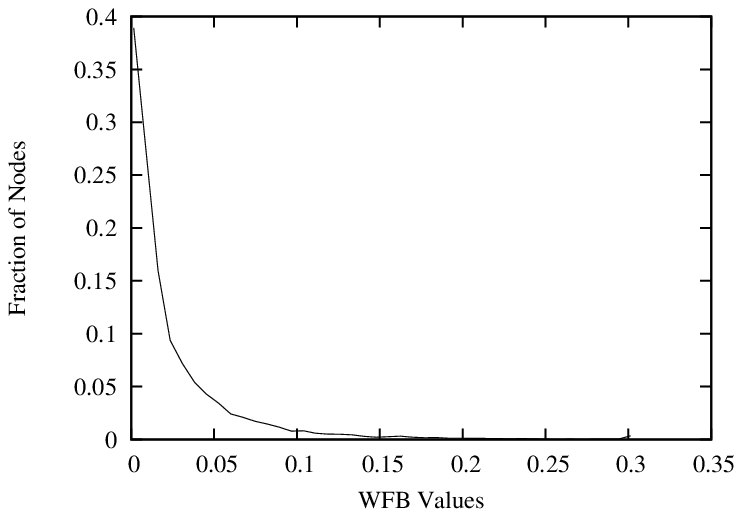}
    \label{subfig:wfbdist_pdf}
    }\quad
    \subfigure[Cumulative distribution of WFB values]{
    \includegraphics[scale=1.0]{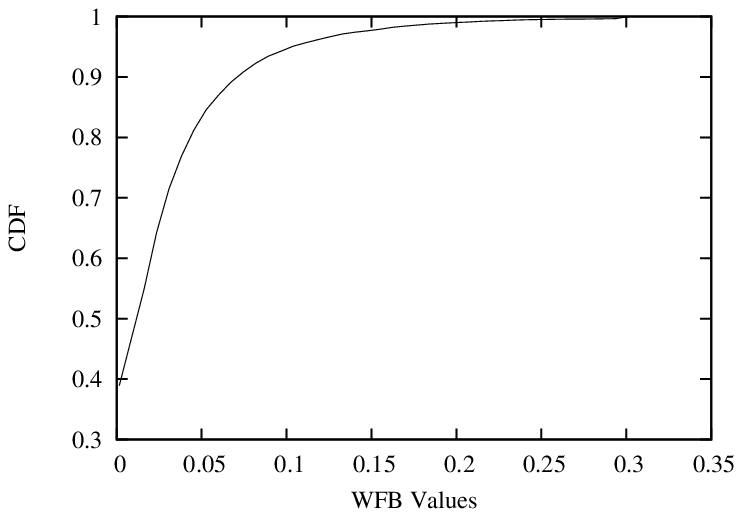}
    \label{subfig:wfbdist_pdf}
    }}

    \caption{Distribution of WFB values in the network.}
    \label{fig:wfb_dist}
\end{figure}

\subsection{Distributed Beamforming Algorithm}\label{subsec:dist-algo}
%We propose an algorithm in which nodes decide on their beamforming behavior only based on their own WFB values and those of their neighbors. 
The distribution of WFB values in the network is shown in Fig. \ref{fig:wfb_dist}. As expected, the majority of nodes have low centrality values while very few nodes have very high values. As with section II-C, we need to identify the set of nodes having the highest values of WFB. A very small fraction of nodes has WFB values in the top 2\%, and can therefore be identified easily. However, we notice that a high fraction of nodes has WFB values lying in the 90-95 percentile. This implies a higher likelihood of these nodes being neighbors and therefore, are likely to result in loss of connectivity if they are all chosen as beamforming nodes. Based on this observation, we propose an algorithm in which nodes decide on their behavior based on WFB values observed in the neighborhood.

As nodes broadcast their WFB values as part of packet transmissions, all nodes are aware of the values of their neighbors. As part of our algorithm, a node $v$ decides on using a directional beam if its own value exceeds that of its neighbors by a \emph{similarity factor} $\beta$. Thus, the beamforming condition for a node $v$ can be expressed as,
\begin{equation}\label{eq:bmf-condn}
\frac{w(v)}{w_{avg}(\mathcal{N}(v))} > \beta
\end{equation}
where $w_{avg}(\mathcal{N}(v))$ denotes the average WFB for $v$'s neighborhood.

The above condition ensures that only nodes with high values of WFB choose themselves for beamforming. The fraction of such nodes is determined by $\beta$. A higher value of $\beta$ implies a stricter condition resulting in fewer beamforming nodes. Further, the condition also results in a lower chance of neighboring nodes creating beams. This is because, a node is chosen if its value exceeds those of its neighbors by the similarity factor $\beta$, the condition is unlikely to hold true for any of the neighbors themselves, and therefore excludes them from beamforming. The choice of beamwidth and beam direction is the same as for centralized choice of beamforming nodes.

\begin{figure}[tb]
    \centering
    \includegraphics[scale=1.0]{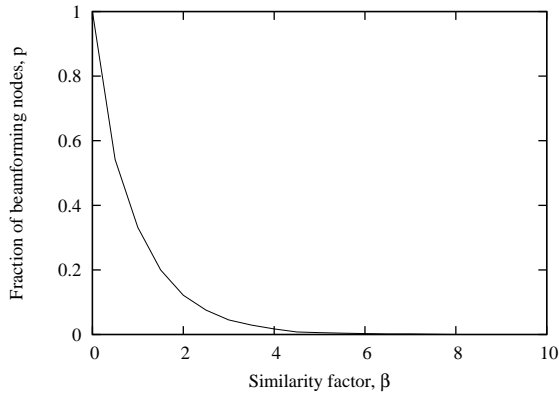}
    \caption{Relation between $\beta$ and $p$}
    \label{fig:pvsbeta}
\end{figure}

The set of nodes that decide on using directional beams is illustrated in Fig. \ref{subfig:nwill_ndchdst} for the network setup in \ref{subfig:nwill_initnw}. Note that, compared to the centralized choice of nodes in Fig. \ref{subfig:nwill_ndchctr}, the distributed choice of nodes are relatively more spread out, thereby reducing the set of neighbors that simultaneously use beams resulting in better connectivity. Figs. \ref{subfig:nwill_ndchdstbms} and \ref{subfig:nwill_ndchdstula} show the set of edges corresponding to the sector and ULA models respectively. The latter results in a greater set of nodes connected using directional beams.

\begin{figure}[tb]
    \centering
\mbox{\subfigure[Path Length Reduction for $d = 4.6$ nodes per unit area]{
    \includegraphics[scale=1.0]{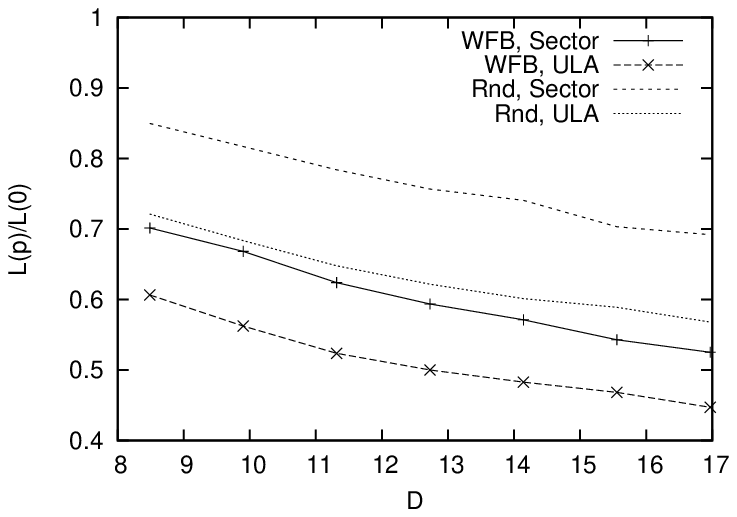}
    \label{subfig:ptlrat_distalgo_d46}
    }\quad
    \subfigure[Path Length Reduction for $d = 3$ nodes per unit area]{
    \includegraphics[scale=1.0]{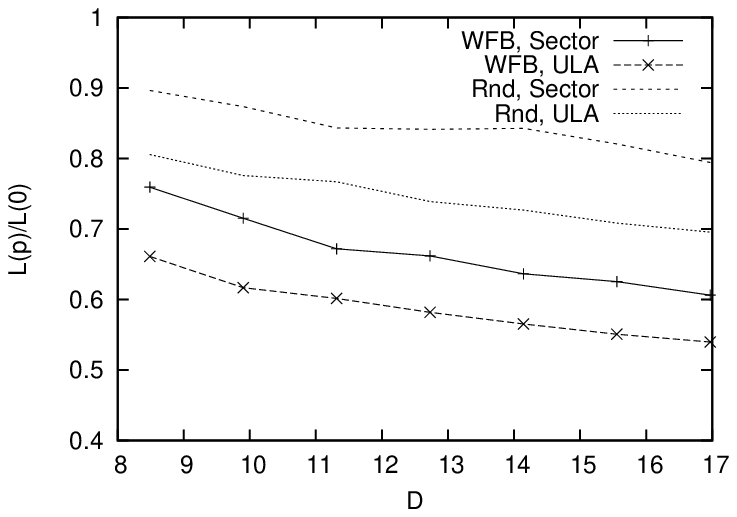}
    \label{subfig:ptlrat_distalgo_d3}
    }}

    \caption{Path length reduction by beamforming for distributed choice of nodes.}
    \label{fig:vardns_distalgo_ptl}
\end{figure}

\begin{figure}[tb]
    \centering
    \includegraphics[scale=1.0]{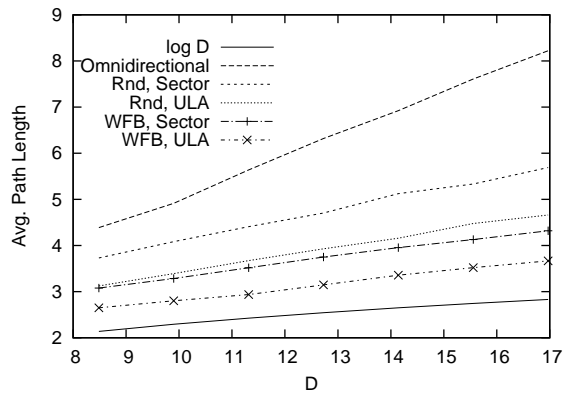}
    \caption{Growth of average path length with the logarithm of the network size.}
    \label{fig:ptgrowth_distalgo}
\end{figure}

\subsection{Simulation Results}\label{subsec:sim-result}
We evaluate the proposed algorithm using simulations at different node densities as earlier. The value of $\beta$ is chosen such that the fraction of beamforming nodes stays close to $p = 0.1$, which was used for the earlier set of simulations using centralized choice of nodes. Using simulations, we identify this value as $\beta = 2$ which results in values of $p$ between $0.11 - 0.13$. The relation between $\beta$ and $p$ is shown in Fig. \ref{fig:pvsbeta}. We obtain results using the sector model as well as the realistic ULA model. As in section \ref{subsec:rndbeam-sim}, the simulation results are obtained using MATLAB. Each individual simulation is averaged over $40$ different topologies and the results obtained as the mean over all possible node pairs in the network.

\begin{figure}[tb]
    \centering
    \mbox{\subfigure[$d = 4.6$]{
    \includegraphics[scale=1.0]{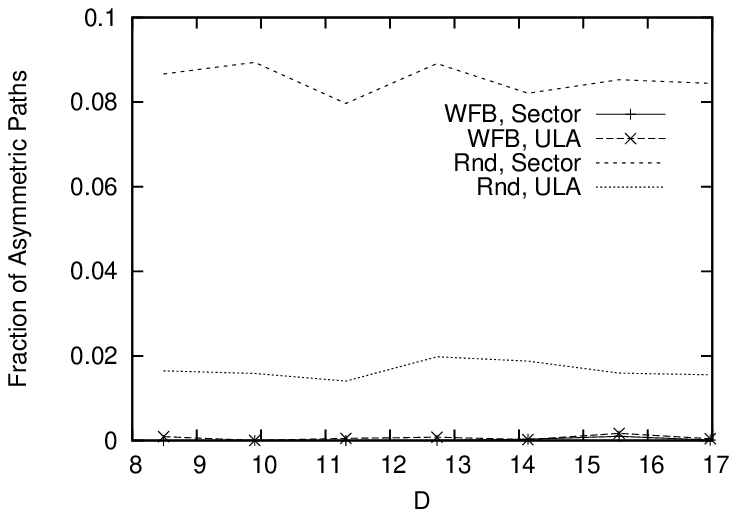}
    \label{subfig:und_distalgo_d46}
    }\quad
    \subfigure[$d = 3$]{
    \includegraphics[scale=1.0]{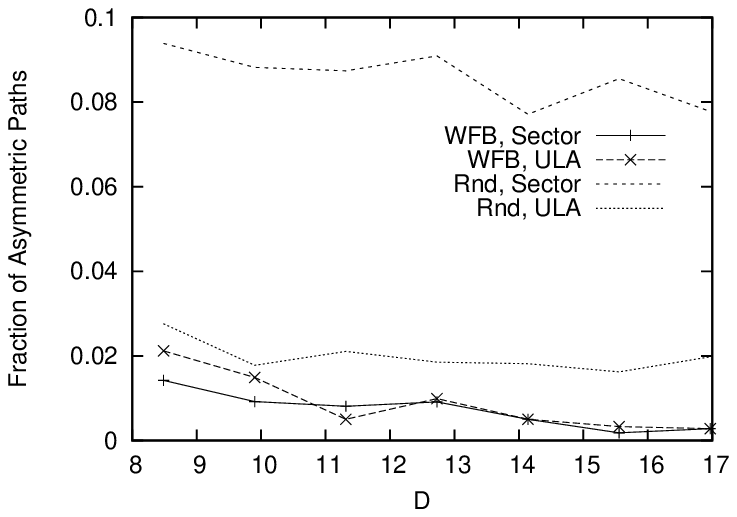}
    \label{subfig:und_distalgo_d3}
    }}
    \caption{Effect on unidirectional connectivity for distributed choice of nodes.}
    \label{fig:vardns_distalgo_und}
\end{figure}
%\begin{figure}[tb]
%    \centering
%\mbox{\subfigure[Unidirectional Connectivity for $d = 4.6$ nodes per unit area]{
%    \includegraphics[scale=1.0]{UniDir_f10_dns4p6_beta2}
%    \label{subfig:unidir_distalgo_d46}
%    }\quad
%    \subfigure[Unidirectional Connectivity for $d = 3$ nodes per unit area]{
%    \includegraphics[scale=1.0]{UniDir_f10_dns3_beta2}
%    \label{subfig:unidir_distalgo_d3}
%    }}
%
%    \caption{Effect on unidirectional connectivity for distributed choice of nodes.}
%    \label{fig:vardns_distalgo_und}
%\end{figure}

Fig. \ref{fig:vardns_distalgo_ptl} illustrates the path length reduction for node densities $d = 3$ and $d = 4.6$ nodes per unit area for both the sector and ULA models. When using the sector model, the path length reduction achieved is higher than in the case of centralized choice of nodes. This additional benefit is again due to the fact that fewer fraction of nodes that are omnidirectional neighbors beamform as a result of the proposed algorithm. Directional beams that are located close to each other tend to reduce the improvement in path length as only one of the two may get chosen most of the time, thereby rendering the other redundant. % Check with visual simulations
The performance for the ULA model is even more promising as greater than $40 \%$ reduction is achieved in both cases, with the improvement being greater than $50 \%$ for greater size of the network region with $d = 4.6$. Fig. \ref{fig:ptgrowth_distalgo} shows the growth in path length for $d = 4.6$ with respect to the logarithmic of the network size. As the choice of beamwidth is the same as earlier, we obtain similar values of $\frac{r(\theta^{*})}{D} \approx 0.2$ for $d = 3$ and $\frac{r(\theta^{*})}{D} \approx 0.25$ for $d = 4.6$ when the network size increases. The reduction in path length, is, therefore, much higher than corresponding values of $\frac{r}{D}$ and $p$ shown in \cite{HelmySWWi}. 

The impact on connectivity is illustrated in Fig. \ref{fig:vardns_distalgo_und}. As in the case of centralized choice of nodes, the fraction of node pairs unidirectionally connected is negligible for higher node density, $d = 4.6$. For lower node density $d = 3$ as well, the impact on connectivity is lower than in the earlier case.
%\begin{figure}[tb]
%    \centering
%    \includegraphics[scale=1.0]{PtLRatWFB_dns4p6_compf_beta2}
%    \caption{Effect on unidirectional connectivity for distributed choice of nodes.}
%    \label{fig:ptlrat_compf}
%\end{figure}

\section{Related Work}\label{sec:rel-work}
\subsection{Self-Organization in Ad Hoc Networks}\label{subsec:litrev-soahn}
Self-organization in wireless networks was analyzed and classified in \cite{DresslerSONAdHoc,PrehoferSlfOrg}. Dressler \cite{DresslerSONAdHoc} classified existing literature on self-organization mechanisms in ad hoc networks based on the information used for distributed decision making. Prehofer and Bettstetter \cite{PrehoferSlfOrg} identified the salient features of a self-organization mechanism. They proposed four design paradigms which form a basis for designing self-organization mechanisms for wireless networks. We discuss how our proposed design for self-organization of wireless networks as small worlds can be associated with the design paradigms in \cite{PrehoferSlfOrg}:
\begin{itemize}
\item \textbf{Local interactions for achieving global properties:} Our design centers around the proposed measure of centrality, the wireless flow betweenness (WFB). As the WFB is computed by nodes using only local neighborhood information, this requirement is satisfied. As shown in section \ref{subsec:comp-wfbfbc}, these local interactions achieve a close correlation to Flow Betweenness Centrality (FBC) computed using global network information.
\item \textbf{Exploit Implicit Coordination:} Implicit coordination in our algorithm results mainly from the use of the neighbor with highest forwarding count for calculation of WFB in equation (\ref{eq:wfb-final}). As such a node is unique within a neighborhood, all nodes have a common understanding of their WFB values.
\item \textbf{Minimize long-lived state information:} As nodes update their own WFB values whenever they overhear a packet transmission in the neighborhood, the state information at nodes is fresh with respect to traffic flows in the network. This design can be enhanced further for dynamically changing topologies such as those considering mobility. Thus, instead of storing WFB values of all neighbors, a node can only store values for those which have recently acted as forwarding nodes. The current discussion is restricted to static networks and hence does not address this aspect.
\item \textbf{Design protocols that adapt to changes:} In the proposed design, both the choice of nodes as well as the beam direction is determined by traffic flows in the network. Thus, it can easily adapt to an increase or decrease in the number of flows in the network.
\end{itemize}

\begin{table}
\centering
%\hspace{-60pt}
\small
\renewcommand{\tabcolsep}{1pt}
\begin{tabular}{|l|l|l|l|l|l|}
\hline
%\multicolumn{2}{|c|}{} & \multicolumn{1}{|c|}{Reference} \\
& \textbf{Parameter/} & \textbf{WFB Based} & \textbf{Guidoni \cite{GuidoniHetSN}} & \textbf{Sharma \cite{SharmaHSN}} & \textbf{Verma \cite{VermaSWWMN}}\\
& \textbf{Algorithm} & & & & \\
\hline
%\multirow{10}{*}{Network model} & Short cut construction & rewiring & addition & addition & addition\\
								& Short cut & Rewiring & Addition & Addition & Addition\\
								& construction & & & & \\
                               \cline{1-6}
%                               & Node distribution & non uniform & uniform & uniform & uniform \\
%                               \cline{2-6}
                               & Node & Single radio & High range & Wired & Two radios\\
                               & Infrastructure & & sensors & & \\
                               \cline{1-6}
                               & Global state & None & Yes & Yes & Yes\\
                               & information & & & & \\
                               \cline{1-6}
%                               & Density of nodes & low & high & high & low\\
%                               \cline{2-6}
                               & Short cut edge & Directed & Undirected & Undirected & Undirected\\
                               \cline{1-6}
                               & Short cut direction & Longest & Random, & Random & Random\\
                               & & recorded path & towards sink & &\\
                               \cline{1-6}
                               & Short cut length & Function of & Constant & Constant & Constant\\
                               & & node density & & &\\
                               \cline{1-6}
%                               & Shortcut Edge width & depends on range & constant & constant & constant\\
%                               \cline{2-6}
                               & p & Function of & Input parameter & Function of & Input parameter\\
                               & & $\beta$ & & network size & \\
                               \cline{1-6}
                               & Performance & Path length & Path length & Path length & Path length\\
                               & metric & Connectivity & Clustering coefficient & Energy dissipation & Clustering coefficient \\
                               & & & & & Avg. Neighbor \\
                               & & & & & degree \\
                               \cline{1-6}
                               \hline
%Performance metric & Connectivity & yes & no & no & no\\
%\hline
\end{tabular}
\caption{Comparison of existing literature on small worlds in wireless networks}
\label{tab:comp-lit}
\end{table}

\subsection{Small Worlds in Wireless Networks}\label{subsec:litrev-swwn}
Small world creation in wireless networks was first investigated by Helmy in \cite{HelmySWWi}. Simulation results were used to study the behaviour of wireless networks as a result of random addition of distance limited short cuts. It was noted in \cite{HelmySWWi} that, owing to the fact that wireless networks are spatial graphs rather than relational, short cut links cannot be completely random as in \cite{WattsStrogSWN}. Rather, the possibility of creating a short cut link between two nodes is determined by the distance between them and the radio transmission range.

Subsequent research has looked into different strategies for creation of short cut links. The authors in \cite{ChitraWired} and \cite{SharmaHSN} considered a hybrid sensor network deployment consisting of a small set of wired links. Theoretical results were obtained in \cite{SharmaHSN} on the relation between the number of such wires required and the average path length. Guidoni et al. in \cite{GuidoniHetSN} showed that small world properties can be achieved in a heterogenous sensor network by using higher capacity nodes to create directed shortcuts towards the sink. In \cite{VermaSWWMN}, Verma et al. proposed three strategies for small world creation in wireless mesh networks in which nodes are equipped with two radios. Short cuts are created between nodes using narrow beam directional links with the additional radios. In \cite{BrustLSWTC,DixitSOCellWiNw}, realization of small world behavior is studied in the context of network scenarios in which a fixed infrastructure is present. The focus, in this case, is topology generation rather than strategies for short cut  creation. Considering the specific case of flows arising from small transactions, Helmy \cite{HelmyContactZone} proposed using border nodes in a multi-hop neighbourhood, defined as the proximity, of a node as shortcuts to optimize route discovery and reduce energy consumption. This, again, differs from the focus of the current work which is to explore shortcut creation to optimize the overall network performance.

The algorithm proposed in this paper is distinguished from the above literature as no additional infrastructure is required to be present for short cut creation. Further, only localized interactions among nodes are made use of instead of using global network information. We compare the main features of our algorithm to those of the existing literature on small worlds in ad hoc networks (identified by surname of the first author and corresponding reference) in Table \ref{tab:comp-lit}. The parameter $p$ denotes the fraction of nodes used for rewiring. Note that we used unidirectional connectivity as a performance metric instead of the clustering coefficient as the former gives a more precise measure of the network performance with regard to the use of directional antennas. Another crucial metric is that of energy dissipation, considered in \cite{SharmaHSN}. In the context of the discussion in this paper, we do not include this as a performance metric since the transmission power for nodes using omnidirectional and directional beams is constant. 

\section{Future Work}\label{sec:fut-work}
As part of our future work, we would like to investigate the navigability of the small world network achieved using the proposed design. As with existing studies on small world networks, decentralized routing algorithms can be designed to identify short paths in the network \cite{KleinbergSWN}. As the small world creation algorithm in this paper makes use of WFB values at nodes, routing can also be designed similarly. As with existing literature in SWNs, such a design needs to minimize the amount of global information available at nodes. Further, we would like to study the use of the WFB measure for other wireless network protocols that can benefit from knowledge of the centrality of nodes.

% Mention about the choice of nodes. Egs. in VANETs, infrastructure nodes have earlier been used to perform certain purposes
While the focus of the current work is to provide overall performance guarantees for the network, it is worth looking at how the proposed algorithm can be adapted to specific network scenarios. In vehicular networks, for instance, existing literature has proposed placing greater burden on fixed infrastructure such as roadside units for data dissemination \cite{SardariRtless,MolinaVisionCoop,YangMHopPeer} . Alternatively, dynamic virtual backbones can be created using the vehicular nodes themselves \cite{BononiCluster,FeliceDynBkbn,DainebiVehClust}. In pocket switched networks, on the other hand, social-network based measure of similarity of nodes plays a role in protocol design \cite{HuiBUBBLERap,DalySNADTN}. Since our definition of WFB is based on information obtained from traffic flows in the network, we expect our design to adapt smoothly to all the above mentioned scenarios. Further, as discussed in section \ref{subsec:litrev-soahn}, the computation of WFB can be adapted depending on the dynamicity of network conditions. A detailed treatment of these issues is out of the scope of the current paper and we leave it for future research.

% Mention about intermittently connected networks. Refer to work on bio-inspired approach
In \cite{AgarwalBioIns}, we investigate the potential of using directional beamforming to realize small-world behaviour in a disconnected ad hoc network. Considering a highly partitioned network, optimal sets of beamforming nodes are identified using bio-inspired algorithms so as to result in a single network component. The primary motivation for this work is to enhance connectivity in sparse networks. This is distinguished from the focus of the current work as here we are motivated to design a self-organization framework that ensures scalability.

\section{Conclusion}\label{sec:concl}
In this paper, we explore the use of directional beamforming for self-organization of a dense wireless ad hoc network as a small world. We provide a simulation based analysis of the achievable performance benefits of randomized beamforming and identify the challenges involved. Subsequently, we detail a distributed algorithm for nodes to decide on their beamforming behavior. We propose a new measure of betweenness centrality, Wireless Flow Betweenness (WFB), which is used to identify beamforming nodes. We show using simulation results that significant performance benefits can be achieved over randomized beamforming.

\end{document}